\documentclass[]{aa}

\usepackage{bm}
\usepackage{amsmath}
\usepackage{xcolor}
\usepackage{graphicx}
\usepackage[outdir=./]{epstopdf}
\usepackage{hyperref}
\usepackage[all]{hypcap}

\begin{document}
	
	\title{Simulations of the Biermann battery mechanism in two-fluid partially ionised plasmas}
	\subtitle{}

	\author{D. Martínez-Gómez\inst{\ref{inst1}}
		\and B. Popescu Braileanu\inst{\ref{inst1},\ref{inst2}}
		\and E. Khomenko\inst{\ref{inst1},\ref{inst2}}
		\and P. Hunana\inst{\ref{inst1}}}
	
	\institute{Instituto de Astrofísica de Canarias, 38205 La Laguna, Tenerife, Spain; \email{dmartinez@iac.es}\label{inst1}
		\and Departamento de Astrofísica, Universidad de La Laguna, 38205 La Laguna, Tenerife, Spain\label{inst2}
	}

	\abstract{In the absence of an initial seed, the Biermann battery term of a non-ideal induction equation acts as a source that generates weak magnetic fields. These fields are then amplified via a dynamo mechanism. The Kelvin-Helmholtz instability is a fluid phenomenon that takes place in many astrophysical scenarios and can trigger the action of the Biermann battery and dynamo processes.}{We aim to investigate the effect that the ionisation degree of the plasma and the interaction between the charged and neutral species have on the generation and amplification of magnetic fields during the different stages of the instability.}{We use the two-fluid model implemented in the numerical code \textsc{Mancha-2F}. We perform 2D simulations starting from a configuration with no initial magnetic field and which is unstable due to a velocity shear. We vary the ionisation degree of the plasma and we analyse the role that the different collisional terms included in the equations of the model play on the evolution of the instability and the generation of magnetic field.}{We find that when no collisional coupling is considered between the two fluids, the effect of the Biermann battery mechanism does not depend on the ionisation degree. However, when elastic collisions are taken into account, the generation of magnetic field is increased as the ionisation degree is reduced. This behaviour is slightly enhanced if the process of charge-exchange is also considered. We also find a dependence on the total density of the plasma related to the dependence on the coupling degree between the two fluids. As the total density is increased, the results from the two-fluid model converge to the predictions of single-fluid models.}{The charged-neutral interaction in a partially ionised plasmas has a non-negligible effect on the Biermann battery mechanism and enhance the generation of magnetic field. In addition, single-fluid models, which assume a very strong coupling between the two species, may overestimate the contribution of this interaction in comparison with two-fluid models.}
	
	\keywords{plasmas; magnetic fields; magnetohydrodynamics (MHD); instabilities}
	
	\titlerunning{Biermann battery and KHI in partially ionised plasmas}
	\authorrunning{}
	
	\maketitle

\section{Introduction} \label{sec:intro}
	Magnetic fields are present in almost every astrophysical system and play a fundamental role in its dynamics. Nevertheless, the issue of their origin is still an open question. A consensus exists that the relatively strong magnetic fields that have been measured in many scenarios, like the solar surface, the interstellar medium or galactic disks, cannot have a primordial origin but are the result of the amplification of a much weaker seed field.
	
	Usually, the amplification process is described within the framework of ideal magnetohydrodynamics (MHD). Magnetic fields become stronger by means of their interaction with the flow of the plasma. Rotational and convective motions within the plasma produce a transformation of kinetic energy into magnetic energy, a phenomenon which is referred to as MHD-dynamo \citep{1919ElecRev..85...412,1933MNRAS..94...39C,1978mfge.book.....M}. However, the induction equation of ideal MHD describes the temporal evolution of a pre-existing magnetic field. That is, it does not allow the creation of new magnetic field when starting from a completely field-free initial condition. Therefore, additional non-ideal processes must be taken into account to explain the origin of magnetic fields.
	
	\citet{1950ZNatA...5...65B} showed that spatial variations of the density and pressure of electrons produce small charge imbalances that may give rise to equally small currents and magnetic fields. The condition for those currents to appear is that the electric field resulting from the charge imbalance must not be curl free, which implies that the gradients of pressure and density must be miss-aligned. This mechanism was in the first place proposed as an explanation for the origin of the magnetic fields of stars \citep[see also][]{1962ApJ...136..615M}. Later, it was applied to other astrophysical contexts like the interestellar medium \citep{1992A&A...264..326L,1997ApJ...480..481K}, the intergalactic medium \citep{1994MNRAS.271L..15S,2000ApJ...539..505G,2008RPPh...71d6901K}, the early universe \citep{2012SSRv..166...37W}, or the Sun's local dynamo \citep{2017A&A...604A..66K}. Descriptions of how a weak magnetic field grows due to the effect of small-scale dynamo can be found in the works of, for example, \citet{1950RSPSA.201..405B} and \citet{2002ApJ...567..828S}.
	
	The Kelvin-Helmholtz instability (KHI) is a common phenomenon in fluid and plasma dynamics. It occurs when perturbations appear at the interface of media moving with different velocities. There is a transfer of energy from the fluid flow to the perturbations so that their amplitudes start to increase. Initially, they follow a phase of exponential growth with a rate that is proportional to the shear flow velocity and inversely proportional to their length scale, as shown by the linear analysis of \citet{1961hhs..book.....C}. Then, during the so-called non-linear stage, the interface becomes greatly distorted due to the formation of large-scale vortexes and as the energy of the flow cascades to smaller scales, so that smaller and smaller vortexes are generated, a turbulent regime is reached \citep{1988FlDyR...3...93K}. The process of inverse cascade in which smaller vortexes merge into larger scale vortexes may be also present in the KHI \citep{2009JA014637}. 
	
	This instability has been observed and studied in a wide variety of astrophysical environments. For instance, in Earth's aurora \citep{1970P&SS...18.1735H}, stellar jets and outflows \citep{1999JPlPh..61....1K}, supernova remnants \citep{2001ApJ...549.1119W}, protoplanetary disks \citep{2005ApJ...630.1093G}, molecular clouds \citep{2007Ap&SS.312...79S}, coronal mass ejections \citep{2011ApJ...729L...8F} and solar spicules \citep{2018ApJ...856...44A}. Its relevance for the research presented in this work resides in the fact that the motions that develop during its evolution allow for the operation of the Biermann battery \citep{2014PhPl...21g2126M} and dynamo \citep{2009ApJ...692L..40Z} mechanisms.
	
	Most of the works previously mentioned regarding the research on the Biermann battery have focussed on the study of fully ionised plasmas. And the ones that have taken into account the effect of partial ionisation have resorted to single-fluid models. The range of applicability of these models is restricted to scenarios in which there is a strong coupling between the charged and neutral components of the plasma. To the best of our knowledge, no previous work has approached the investigation of the Biermann mechanism in partially ionised plasmas by using multi-fluid models, which are expected to have a broader range of applicability. On the other hand, the multi-fluid framework has already been used for the study of the KHI in partially ionised plasmas, as shown in \citet{2004ApJ...608..274W}, \citet{2012ApJ...749..163S}, \citet{2015AA...578A.104M} or \citet{2019PhPl...26h2902H}. However, those works did not include the battery term in their models.
	
	Therefore, the goal of this article is to try to fill the above mentioned gap and improve the understanding of the magnetogenesis processes in partially ionised plasmas. For that purpose, we use a two-fluid model that treats the neutral component of the plasma as a separate fluid from the charged one (which includes ions and electrons) but allows them to interact with each other by means of collisions. Here, we focus on fluid models with isotropic and scalar pressures. For a review of so-called collisionless fluid models with anisotropic pressures (temperatures), see \citet{2019JPlPh..85f2003H,2019JPlPh..85f2002H}. Our model uses a generalised Ohm's law \citep{2014PhPl...21i2901K} and takes into consideration several collisional terms, such as the momentum and heat transfer due to elastic collisions and charge-exchange collisions in the momentum and energy equations, and a collisional term in the induction equation. Due to the complexity of the dynamics of the system we are interested in, we find the necessity of basing our research on numerical simulations. Hence, the equations of our model are solved using the code \textsc{Mancha-2F} \citep{2019A&A...627A..25P}. Nevertheless, we also present an analysis of the equations, which helps in understanding the numerical results. In addition, we perform a brief comparison between our numerical results and the predictions from single-fluid models.
	
	For the time being, we restrict our simulations to the 2D case and use plasma parameters that correspond to the solar chromosphere. However, we expect that our results are general enough for their application to other partially ionised plasmas and that they may be easily extrapolated to other astrophysical scenarios.

	The present paper is organised as follows. In Sect. \ref{sec:method}, we show the equations of the two-fluid model and describe the numerical setup for the simulation of the KHI. In Sect. \ref{sec:sims}, we present the results of several series of simulations. We describe the evolution of the instability and the growth of the magnetic field. Then, we study the dependence on the ionisation degree of the plasma, on the different collisions terms included in the model and on the numerical dissipation. In Sect. \ref{sec:disc}, we analyse the equations of our model to explain the results obtained from the simulations. Finally, in Sect. \ref{sec:conc}, we present a summary of the performed research and discuss some lines that can be followed to improve it.

\section{Numerical model} \label{sec:method}
	In this work, we use a two-fluid model to describe the properties and the temporal evolution of a partially ionised plasma. The plasma is assumed to be composed of three different kinds of particles: ions, electrons and neutrals, which are denoted by the subscripts ``i'', ``e'' and ``n'', respectively. However, a strong coupling between ions and electrons is assumed and they are taken into account together as a charged (or ionised) fluid denoted by the subscript ``c''. The other fluid is composed of neutral particles only. The derivation of this two-fluid model from a more general multi-fluid description can be found in \citet{2014PhPl...21i2901K}.
	
	The equations that govern the evolution of the density, velocity and energy of each fluid and of the magnetic field are shown in Sect. \ref{sec:eqs}. These equations are solved numerically using the \textsc{Mancha-2F} code, which is a two-fluid extension of the original single-fluid \textsc{Mancha3D} code \citep{2006ApJ...653..739K,2010ApJ...719..357F,2018A&A...615A..67G}. The details of the numerical implementation of the two-fluid model can be found in \cite{2019A&A...627A..25P}. The specific setup of initial and boundary conditions used in the present work is explained in Sect. \ref{sec:setup}.
	 
\subsection{Equations} \label{sec:eqs}
    The evolution of the density of each one of the two fluids is described by its respective continuity equation:
    \begin{equation} \label{eq:rho}
		\frac{\partial \rho_{\rm{s}}}{\partial t} + \nabla \cdot \left(\rho_{\rm{s}} \bm{V}_{\rm{s}}\right) = 0,
	\end{equation}
    where $s \in \{c,n\}$, and $\rho_{\rm{c}} = \rho_{\rm{i}} + \rho_{\rm{e}}$. No source terms appear on the right-hand side because we do not consider the processes of ionisation and recombination.
    
	Neglecting the electrons inertia due to their small mass, $m_{\rm{e}} \ll m_{\rm{i}}$, the velocity of the charged fluid is given by the velocity of the ions, that is $\bm{V}_{\rm{c}} = \bm{V}_{\rm{i}}$. Neglecting also the effect of gravity, the velocities $\bm{V}_{\rm{c}}$ and $\bm{V}_{\rm{n}}$ are computed from the following momentum equations:
	\begin{equation} \label{eq:momc}
		\frac{\partial \left(\rho_{\rm{c}} \bm{V}_{\rm{c}}\right)}{\partial t} + \nabla \cdot \left( \rho_{\rm{c}} \bm{V}_{\rm{c}} \bm{V}_{\rm{c}} + P_{\rm{c}} \mathbb{I} \right) = \bm{J} \times \bm{B} + \bm{R}_{\rm{cn}}
	\end{equation}
	and
	\begin{equation} \label{eq:momn}
		\frac{\partial \left(\rho_{\rm{n}} \bm{V}_{\rm{n}}\right)}{\partial t} + \nabla \cdot \left( \rho_{\rm{n}} \bm{V}_{\rm{n}} \bm{V}_{\rm{n}} + P_{\rm{n}} \mathbb{I} \right) = \bm{R}_{\rm{nc}}.
	\end{equation}
    The variable $P_{\rm{c}}$ represents the sum of the pressures of ions and electrons, $P_{\rm{n}}$ is the pressure of neutrals, $\mathbb{I}$ is the identity tensor, $\bm{J}$ is the current density and $\bm{B}$ is the magnetic field. Neglecting the displacement currents in Ampère's law, the variables $\bm{J}$ and $\bm{B}$ are related by:
    \begin{equation}
	    \bm{J} = \frac{\nabla \times \bm{B}}{\mu_{0}},
    \end{equation}
    where $\mu_{0}$ is the magnetic permeability.
    
    The terms $\bm{R}_{\rm{cn}}$ and $\bm{R}_{\rm{nc}}$ represent the momentum transfer due to collisions between two fluids moving with different velocities and are defined as
    \begin{equation} \label{eq:Rcoll}
    	\bm{R}_{\rm{cn}} = \alpha_{\rm{eff}} \rho_{\rm{c}} \rho_{\rm{n}} \left(\bm{V}_{\rm{n}} - \bm{V}_{\rm{c}} \right) = -\bm{R}_{\rm{nc}}.
    \end{equation}
    The collisional parameter $\alpha_{\rm{eff}}$ takes into account the elastic collisions between ions and neutrals, between electrons and neutrals, and the charge-exchange collisions between ions and neutrals. Thus,
    \begin{equation}
    	\alpha_{\rm{eff}} = \alpha + \alpha_{\rm{cx}},
    \end{equation}
    where $\alpha$ is the contribution of the elastic collisions and $\alpha_{\rm{cx}}$ is the contribution of the charge-exchange collisions.
    
    The elastic collisional parameter is given by
    \begin{equation} \label{eq:alpha_def}
    	\alpha = \frac{\rho_{\rm{e}} \nu_{\rm{en}} + \rho_{\rm{i}} \nu_{\rm{in}}}{\rho_{\rm{n}} \rho_{\rm{c}}},
    \end{equation}
    where $\nu_{\rm{st}}$ is the frequency of collisions of particles ``s'' with particles ``t''. Following the works of \citet{1965RvPP....1..205B} and \citet{1986MNRAS.220..133D}, these frequencies are computed as
    \begin{equation} \label{eq:nu_st}
    	\nu_{\rm{st}} = \frac{n_{\rm{t}} m_{\rm{t}}}{m_{\rm{s}}+m_{\rm{t}}} \sqrt{\frac{8 k_{\rm{B}}}{\pi}\left(\frac{T_{\rm{s}}}{m_{\rm{s}}}+\frac{T_{\rm{t}}}{m_{\rm{t}}}\right)} \sigma_{\rm{st}},
    \end{equation} 
    where $m_{\rm{s}}$ is the mass of a particle ``s'', $T_{\rm{s}}$ is the temperature, $n_{\rm{t}}$ is the number density of species ``t'', $n_{\rm{t}}=\rho_{\rm{t}} / m_{\rm{t}}$, $k_{\rm{B}}$ is Boltzmann constant and $\sigma_{\rm{st}}$ is the collisional cross-section. Then, the parameter $\alpha$ can be computed through the following expression
    \begin{equation} \label{eq:alpha}
    	\alpha = \frac{m_{\rm{in}}}{m_{\rm{n}}^{2}} \sqrt{\frac{8k_{\rm{B}}}{\pi}\frac{T_{\rm{cn}}}{m_{\rm{in}}}} \sigma_{\rm{in}} + \frac{m_{\rm{en}}}{m_{\rm{n}}^{2}} \sqrt{\frac{8k_{\rm{B}}}{\pi}\frac{T_{cn}}{m_{\rm{en}}}} \sigma_{\rm{en}},
    \end{equation}
    in which $T_{\rm{cn}} = \left(T_{\rm{c}}+T_{\rm{n}}\right) / 2$ is the average temperature and $m_{\rm{st}} = m_{\rm{s}} m_{\rm{t}} / \left(m_{\rm{s}} + m_{\rm{t}}\right)$ is the reduced mass of particles ``s'' and ``t''. For the sake of simplicity, in the present study we assume that the plasma is composed of hydrogen only. Therefore, $m_{\rm{i}} \approx m_{\rm{n}}$ is the proton mass, $m_{\rm{in}} =0.5 m_{\rm{i}}$ and $m_{\rm{en}} \approx m_{\rm{e}}$. We choose the following values of the collisional cross sections \citep{2013PhPl...20f1202L}: $\sigma_{\rm{in}} = 1.16 \times 10^{-18} \ \rm{m^{2}}$ and $\sigma_{\rm{en}} = 10^{-19} \ \rm{m^{2}}$. Then, $\alpha$ can be expressed as
    \begin{equation} \label{eq:alpha_2}
        \alpha = \frac1{m_{n}^{2}}\sqrt{\frac{8 k_{\rm{B}}}{\pi}T_{\rm{cn}}}\left[\sqrt{0.5m_{\rm{i}}}\sigma_{\rm{in}}+\sqrt{m_{\rm{e}}}\sigma_{\rm{en}}\right],
    \end{equation}
    and since $m_{\rm{e}} \ll m_{\rm{i}}$ and $\sigma_{\rm{en}} < \sigma_{\rm{in}}$, the value of this parameter is mainly determined by the ion-neutral interaction, with a much smaller contribution coming from the electron-neutral collisions.
    
    According to the definition of collision frequencies given by Eq. (\ref{eq:nu_st}), the consevation of momentum is satisfied:
    \begin{equation}
    	\rho_{\rm{s}} \nu_{\rm{st}} = \rho_{\rm{t}} \nu_{\rm{ts}}.
    \end{equation}
    In addition, the expression for the collisional parameter $\alpha$ given by Eq. (\ref{eq:alpha_def}) allows us to define effective collision frequencies between charges and neutrals as
    \begin{equation} \label{eq:nu_cn_nc} 
    	\nu_{\rm{cn}} = \alpha \rho_{\rm{n}}  \ \text{and} \ \nu_{\rm{nc}} = \alpha \rho_{\rm{c}}.
    \end{equation}
    
    The contribution of the charge-exchange collisions is computed as (see, e.g., \citet{1995JGR...10021595P,2012PhPl...19g2508M}):
    \begin{equation} \label{eq:alpha_cx}
    	\alpha_{\rm{cx}} = \frac1{m_{\rm{n}}} \left(V_{0}^{\rm{cx}} + V_{\rm{cn}}^{\rm{cx}}+V_{\rm{nc}}^{\rm{cx}} \right) \sigma_{\rm{cx}},
    \end{equation}
    where the velocities $V_{0}^{\rm{cx}}$ and $V_{\rm{st}}^{\rm{cx}}$ are defined as
    \begin{equation}
    	V_{0}^{\rm{cx}} = \sqrt{\frac{4}{\pi}v_{\rm{Tc}}^{2} + \frac{4}{\pi}v_{\rm{Tn}}^{2}+v_{\rm{D}}^{2}},
    \end{equation}
    \begin{equation}
	    V_{\rm{st}}^{\rm{cx}} = v_{\rm{Ts}}^{2} \left[4 \left(\frac{4}{\pi}v_{\rm{Tt}}^{2}+v_{\rm{D}}^{2} \right)+\frac{9 \pi}{4}v_{\rm{T s}}^{2} \right]^{-0.5}
    \end{equation}
    and the cross section $\sigma_{\rm{cx}}$ is given by
    \begin{equation}
	    \sigma_{\rm{cx}} = 1.12 \times 10^{-18} - 7.15 \times 10^{-20} \ln \left(V_{0}^{\rm{cx}} \right).
    \end{equation}
    The thermal speed of each species ``s'' is given by
    \begin{equation} \label{eq:vtherm}
    	v_{\rm{Ts}} = \sqrt{\frac{2k_{\rm{B}}T_{\rm{s}}}{m_{\rm{s}}}},
    \end{equation}
    and
    \begin{equation}
    	v_{\rm{D}} = |\bm{V}_{\rm{c}}-\bm{V}_{\rm{n}}|
    \end{equation}
    is the velocity drift between charged and neutral particles.
    
    The evolution of the magnetic field is given by the induction equation, namely
    \begin{equation} \label{eq:induction}
    	\frac{\partial \bm{B}}{\partial t} = -\nabla \times \bm{E},
    \end{equation}
    where the electric field $\bm{E}$ is given by a generalised Ohm's law. Derivations of Ohm's law in different multi-fluid descriptions can be found in, for instance, \citet{2011A&A...529A..82Z}, \citet{2016ApJ...832..101M} and \citet{2018SSRv..214...58B}. Here we follow the approach used by \citet{1986fopp.book.....B} and \citet{2014PhPl...21i2901K}, with the difference that we neglect the effect of resistivity and Hall's term. Therefore, the Ohm's law used in the present work is given by
    \begin{equation} \label{eq:ohm}
    	\bm{E} = -\bm{V_{\rm{c}}} \times \bm{B} - \eta_{\rm{H}} \nabla P_{e} + \eta_{\rm{D}} \left(\bm{V}_{\rm{n}} - \bm{V}_{\rm{c}} \right),
    \end{equation}
	where
	\begin{equation} \label{eq:eta_h}
		\eta_{\rm{H}} = \frac1{e n_{\rm{e}}},
	\end{equation}
	
	\begin{equation}
		\eta_{\rm{D}} = \frac{m_{\rm{e}} \left(\nu_{\rm{en}} - \nu_{\rm{in}} \right)}{e},
	\end{equation}
	and $e$ is the elemental charge.
	
	Combining Eqs. (\ref{eq:induction}) and (\ref{eq:ohm}) we get the following expression for the induction equation:
	\begin{eqnarray} \label{eq:induction2}
		\frac{\partial \bm{B}}{\partial t} &=& \nabla \times \left( \bm{V}_{\rm{c}} \times \bm{B} \right) + \nabla \times \left[\eta_{H} \nabla P_{\rm{e}} \right] \nonumber \\
		&-& \nabla \times \left[\eta_{\rm{D}} \left(\bm{V}_{\rm{n}} - \bm{V}_{\rm{c}} \right)\right] = \bm{A}_{\rm{Ind}} + \bm{B}_{\rm{Ind}} + \bm{C}_{\rm{Ind}},
	\end{eqnarray}
	where $\bm{A}_{\rm{ind}}$, $\bm{B}_{\rm{Ind}}$ and $\bm{C}_{\rm{Ind}}$ are the advective, Biermann battery and collisional terms, respectively. The collisional term is usually neglected because in the presence of a background magnetic field it is typically very small in comparison with the advection term. However, in this work we are interested in a situation in which there is no initial magnetic field. So, we keep this term to check whether in this case it may have a relevant influence on the evolution of the system.
	
	It is interesting to consider the scenario with no initial magnetic field and no collisions between the charged and the neutral fluids. In this case, the only term that appears on the right-hand side of the induction equation is the Biermann battery term. Using the definition given by Eq. (\ref{eq:eta_h}) for the coefficient $\eta_{H}$, the induction equation has the following form:
	\begin{equation} \label{eq:indu_batt}
  		\frac{\partial \bm{B}}{\partial t} = \nabla \times \left[ \frac{\nabla P_{\rm{e}}}{e n_{\rm{e}}} \right] = -\frac{\nabla n_{\rm{e}} \times \nabla P_{\rm{e}}}{e n_{\rm{e}}^{2}}.
  	\end{equation}
  	
  	This formula tells us that, even when there is not an initial magnetic field seed, a magnetic field is generated if the spatial variations in the density and pressure of electrons are misaligned. It must be noted here that this effect will not be present if the flow is barotropic, that is, if the pressure is a function of the density only, $P_{e} = P_{e} (n_{e})$ \citep{1997ApJ...480..481K}.
	
	The equations for the evolution of the internal energy of each fluid, $e_{\rm{c}}$ and $e_{\rm{n}}$, respectively, are
	\begin{equation} \label{eq:iener_c}
		\frac{\partial e_{\rm{c}}}{\partial t}+\nabla \cdot \left(e_{\rm{c}}\bm{V}_{\rm{c}} \right)+P_{c}\nabla \cdot \bm{V}_{\rm{c}}=\bm{J} \cdot \bm{E}_{\rm{diff}} + Q_{\rm{cn}}
	\end{equation}
	and
	\begin{equation} \label{eq:iener_n}
		\frac{\partial e_{\rm{n}}}{\partial t} + \nabla \cdot \left(e_{\rm{n}}\bm{V}_{\rm{n}} \right)+P_{\rm{n}} \nabla \cdot \bm{V}_{\rm{n}} = Q_{\rm{nc}},
	\end{equation}
	where
	\begin{equation} \label{eq:ediff}
		\bm{E}_{\rm{diff}}=\bm{E}+\bm{V}_{\rm{c}} \times \bm{B}
	\end{equation}
	and $Q_{\rm{cn}}$ and $Q_{\rm{nc}}$ are the heat transfer terms associated with collisions.
	
	The internal energies are related to the pressures by
	\begin{equation} \label{eq:iene_pres}
		e_{\rm{c}}=\frac{P_{\rm{c}}}{\gamma-1} \ \ \rm{and} \ \ e_{n}=\frac{P_{n}}{\gamma-1},
	\end{equation}
	and, since $m_{\rm{i}} \approx m_{\rm{n}}$, the energy transfer terms due to collisions \citep[see, e.g,][]{1977RvGSP..15..429S,1986MNRAS.220..133D}  are given by

	\begin{equation} \label{eq:qcn}
		Q_{\rm{cn}} = \alpha_{\rm{eff}} \rho_{\rm{c}}\rho_{\rm{n}}\left[\frac{k_{\rm{B}}}{\left(\gamma-1\right)m_{\rm{n}}}\left(T_{\rm{n}}-T_{\rm{c}}\right)+\frac{v_{\rm{D}}^{2}}{2}\right]
	\end{equation}
	and
	\begin{equation} \label{eq:qnc}
		Q_{\rm{nc}} = \alpha_{\rm{eff}} \rho_{\rm{c}}\rho_{\rm{n}}\left[\frac{k_{\rm{B}}}{\left(\gamma-1\right)m_{\rm{n}}}\left(T_{\rm{c}}-T_{\rm{n}}\right)+\frac{v_{\rm{D}}^{2}}{2}\right].
	\end{equation}
	
	We note here that the expressions used for the momentum and energy transfer due to collisions, that is, Eqs. (\ref{eq:Rcoll}), (\ref{eq:qcn}) and (\ref{eq:qnc}) are only valid when the velocity drifts between the two species are much smaller than the thermal speeds of the fluids. The general expressions include correction factors that depend on the ratios between those speeds \citep{1977RvGSP..15..429S,1986MNRAS.220..133D}.
	
	Finally, as equations of state we use the ideal gas law, which relates the pressure, number density and temperature of a fluid in the following way:
	\begin{equation} \label{eq:state}
		P_{\rm{s}} = n_{\rm{s}} k_{\rm{B}} T_{\rm{s}}.
	\end{equation}
	Due to the assumptions of thermal equilibrium between ions and electrons ($T_{\rm{i}} = T_{\rm{e}} = T_{\rm{c}}$) and of charge neutrality of the plasma ($n_{\rm{i}} = n_{\rm{e}}$), we have that
	\begin{equation} \label{eq:state_2}
		P_{\rm{c}} = n_{\rm{c}} k_{\rm{B}} T_{\rm{c}}= 2 n_{\rm{i}} k_{\rm{B}} T_{\rm{c}} = 2 P_{\rm{i}} = 2 P_{\rm{e}}.
	\end{equation}
	
	From the set of equations presented in the previous paragraphs, it can be checked that the only physical dissipative mechanism taken into account is the friction due to collisions between charged and neutral particles. This model does not include other real dissipative processes such as viscosity or resistivity (which, in fact, are different kinds of collisional interactions). 
	
	However, the discretisation of these equations in order to solve them numerically leads to the appearance of high-frequency numerical noise. To reduce this unwanted noise and assure the stability of the simulations, a numerical diffusivity term is added to the right-hand side of each of the evolution equations. Details on the implementation of these terms in the \textsc{Mancha3D} and \textsc{Mancha-2F} codes can be found in \citet{2005A&A...429..335V} and \citet{2010ApJ...719..357F}. In general, each numerical diffusivity term is composed of a shock-resolving term, a hyperdiffusivity part and a constant contribution. In this work, we only consider the latter. 
	
	The numerical diffusion terms act as analogues of the viscous forces and the Ohmic dissipation and have a non-negligible impact on the dynamics of the plasma. In Appendix \ref{sec:sim_diss}, we present a brief study of the influence of the magnitude of the numerical dissipation on the results of our simulations.
	
\subsection{Numerical setup} \label{sec:setup}
	We consider a physical rectangular domain of dimensions $\left[-L_{0}/2, L_{0}/2 \right] \times [-L_{0},L_{0}]$, where $L_{0}$ is the length in the $x$-direction, represented by a numerical mesh of $N_{0} \times 2 N_{0}$ points. Periodic boundary conditions are applied in both directions. 
	
	This region is filled with a partially ionised plasma whose density is uniform along the horizontal direction but varies in the vertical direction. More precisely, the vertical density profile corresponds to that of a denser slab embedded in a lighter medium, with smooth transitions between the internal and external layers. For the plasma to be in mechanical equilibrium, the pressure of both the charged and the neutral fluids must be uniform. This implies that the initial temperature varies with height in a way that is inversely proportional to the density. 
	
	The ionisation degree of the plasma is defined as $\chi_{\rm{c}} = \rho_{\rm{c}} / \rho_{\rm{T}}$, where $\rho_{\rm{T}} = \rho_{\rm{c}} + \rho_{\rm{n}}$ is the total density. In the same way, the density fraction of neutrals is given by $\chi_{\rm{n}} = \rho_{\rm{n}} / \rho_{\rm{T}} = 1 - \chi_{c}$. In general these parameters vary with the density, pressure and temperature of the fluids. However, for the sake of simplicity, here we neglect this dependence and assume that $\chi_{\rm{c}}$ and $\chi_{\rm{n}}$ are uniform at the initial time.

    Then, the vertical density profile of each fluid ``s'' is given by the following expression:
    \begin{eqnarray} \label{eq:ini_rho}
    	\lefteqn{\rho_{\rm{s}} (z, t= 0)= \frac{\chi_{\rm{s}}}{2} \Bigg[ 2\rho_{\rm{T1}} + \left(\rho_{\rm{T2}} - \rho_{\rm{T1}}\right) \tanh \left(\frac{z - z_{\rm{L1}}}{w_{\rm{L}}} \right)} \nonumber \\
    	& & + \left(\rho_{\rm{T1}} - \rho_{\rm{T2}}\right) \tanh \left(\frac{z - z_{\rm{L2}}}{w_{\rm{L}}} \right) \Bigg],
    \end{eqnarray}
	where $\rho_{\rm{T1}}$ is the total density of the top and bottom regions, $\rho_{\rm{T2}}$ is the total density of the central part, $z_{\rm{L1}}$ and $z_{\rm{L2}}$ are the central positions of the transition layers, and $w_{\rm{L}}$ is the width of those layers. For our simulations, we have set the following values of these parameters: $\rho_{\rm{T2}} = 2 \rho_{\rm{T1}}$, $z_{\rm{L1}} = -L_{0}/2$, $z_{\rm{L2}} = L_{0} / 2$, and $w_{\rm{L}} = L_{0} / 25$.
	
	As already mentioned, the KHI appears when there is a shear flow at the interface of two media. Therefore, to simulate this instability we consider an initial velocity profile in which the fluids at the central heavier region are moving along the $x$-direction at a speed $U_{\rm{2,s}}$, while the top and bottom regions move at a speed $U_{\rm{1,s}}$, with a smooth variation between those two values in the transition layers. 
	
	Since the evolution of the KHI depends on the velocity drift between the different media, defined as $\Delta U \equiv |U_{\rm{1,s}}-U_{\rm{2,s}}|$, we can move to a reference frame in which one of the media is at rest. Here, we have chosen the outer layers to be at rest, so $U_{\rm{1,s}} = 0$, while the central region is moving at a speed $U_{\rm{2,s}} = U_{0}$. Therefore, the initial flow along the $x$-direction is given by
	\begin{equation} \label{eq:ini_vx2}
		V_{x,\rm{s}} (z,t=0) = \frac{U_{0}}{2} \Bigg[\tanh \left(\frac{z - z_{\rm{L1}}}{w_{\rm{L}}} \right) - \tanh \left(\frac{z-z_{\rm{L2}}}{w_{\rm{L}}} \right) \Bigg].
	\end{equation} 
	
	Then, we apply a small perturbation to the $z$-component of the velocity, given by
	\begin{eqnarray} \label{eq:ini_vz}
		\lefteqn{V_{z,\rm{s}}(x,z,t=0)=-V_{0} \sin \left(\frac{2\pi}{L_{0}}  x\right) \exp \left(-\frac{|z-z_{\rm{L1}}|}{0.5 w_{\rm{L}}} \right)} \nonumber \\
		& & + V_{0} \sin \left(\frac{2\pi}{L_{0}}x \right) \exp \left(-\frac{|z-z_{\rm{L2}}|}{0.5 w_{\rm{L}}}\right),
	\end{eqnarray}
	where the value of $V_{0}$ is much smaller than the maximum flow speed of the plasma in the $x$-direction, that is $V_{0} \ll U_{0}$. This perturbation corresponds to an oscillation of wavelength $\lambda = L_{0}$, with a maximum amplitude at the position of the transition layers and exponentially tends to zero when going away from those positions.
	
\begin{figure}
		\centering
		\resizebox{\hsize}{!}{\includegraphics[]{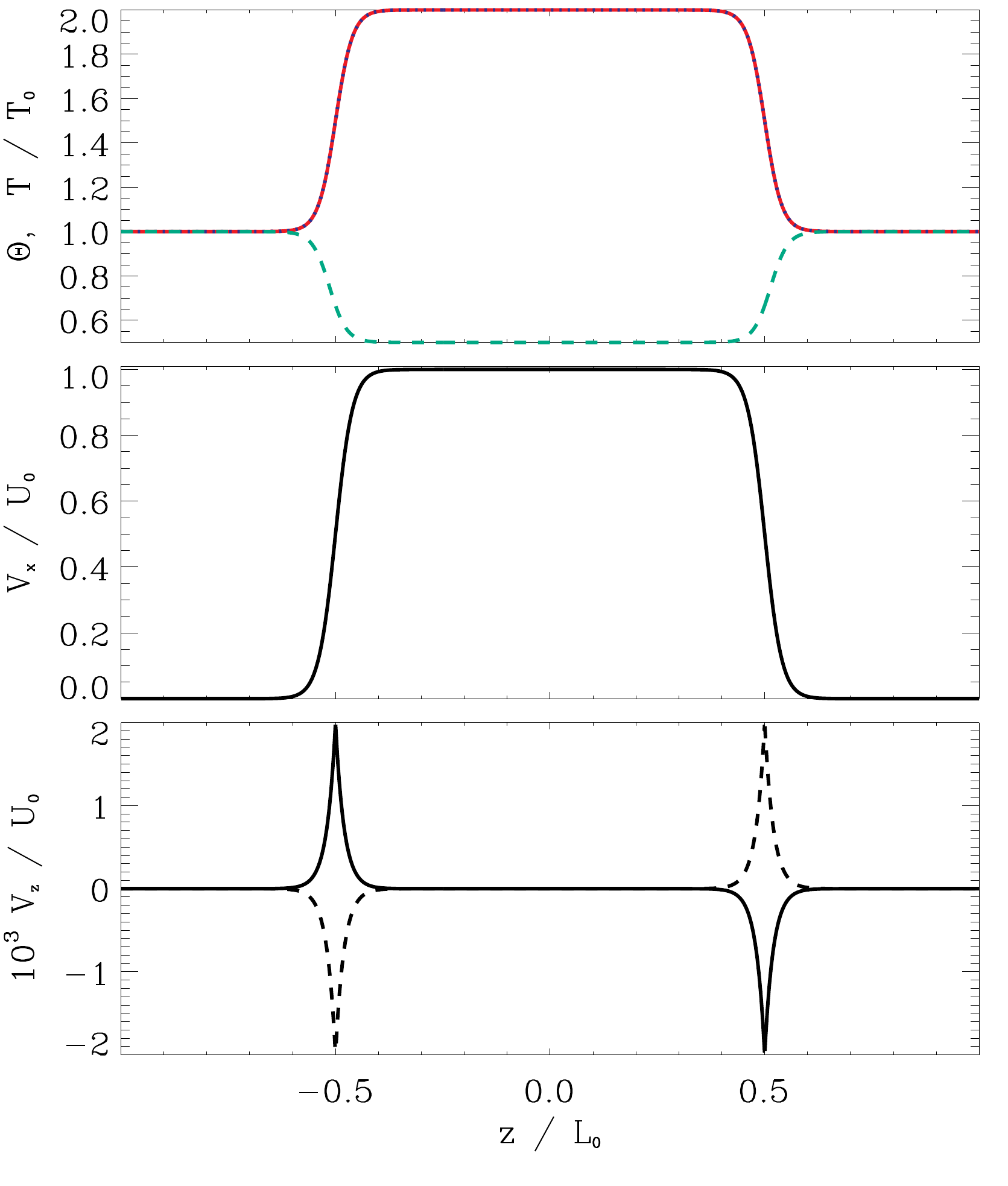}}
		\caption{Vertical cuts of the initial conditions of the simulations. Top: normalised density, $\Theta$ (red solid line) and normalised temperature (green dashed line) at $x = 0$. Middle: normalised flow speed at $x = 0$. Bottom: normalised perturbation of the $z$-component of the velocity at $x = -L_{0} / 4$ (solid line) and $x = L_{0} /4$ (dashed line).}
		\label{fig:ini_conds}
	\end{figure}

	Figure \ref{fig:ini_conds} shows vertical cuts of the initial conditions described in the paragraphs above. The top panel shows the normalised temperature, $T / T_{0}$ (where $T_{0}$ is the temperature of the outer regions), as a dashed green line and the normalised profile of density, $\Theta$, as a solid red line. This parameter is computed as $\Theta= \rho_{\rm{s}} (z) / (\chi_{s} \rho_{\rm{T1}})$. The vertical profile of the normalised velocity in the $x$-direction, $V_{x} / U_{0}$, is represented at the middle panel. The normalised initial perturbation in the $z$-component of the velocity, $V_{z} / U_{0}$ is shown at the bottom panel: the solid line corresponds to the position $x = -L_{0} / 4$, while the dashed line corresponds to $x = L_{0} / 4$.

	Finally, since we are interested in studying how partial ionisation affects the generation of magnetic field by means of the Biermann battery effect, we neglect the presence of an initial background magnetic field.
	
	We note that with the setup described in the previous paragraphs, the pressure of each species ``s'' is not only a function of the fluid density but it also depends on the temperature, which is non-uniform. It does not fulfill that $P_{\rm{s}} = P_{\rm{s}}(\rho_{\rm{s}})$ but $P_{\rm{s}} = P_{\rm{s}} (\rho_{\rm{s}}, T_{\rm{s}})$. Therefore, this configuration allows the action of the Biermann battery effect.
	
\section{Results of simulations} \label{sec:sims}
	For the present study, we have used values of the initial densities and temperatures typically found in the solar chromosphere. It has been shown by, for instance, \citet{2015ApJ...813..123Z}, \citet{2018ApJ...856...44A} and \citet{2019SoPh..294...20Z} that motions in chromospheric structures, such as spicules, may lead to the development of the KHI. Using those structures as references, the initial total density and temperature in our simulations are $\rho_{\rm{T1}} = 10^{-10} \ \rm{kg \ m^{-3}}$ and $T_{0} = 10^{4} \ \rm{K}$. Since the sound speed of each fluid ``s'' is
	\begin{equation} \label{eq:csc}
		c_{\rm{S,s}} = \sqrt{\frac{\gamma P_{\rm{s}}}{\rho_{\rm{s}}}},
	\end{equation}
	using Eqs. (\ref{eq:state}) and (\ref{eq:state_2}), the chosen temperature gives values of $c_{\rm{S,c}} \approx 16.6 \ \rm{km \ s^{-1}}$ and $c_{\rm{S,n}} \approx 11.7 \ \rm{km \ s^{-1}}$.

	In addition, we have set the length-scale to be $L_{0} = 10^{4} \ \rm{m}$. This length is small in comparison with the thickness or height of the spicules but it is an appropriate value for the investigation of the effects of the charged-neutral interaction.
	
	The works of, for example, \citet{2005A&A...442.1091L}, \citet{2011A&A...529A..82Z}, \citet{2013ApJ...767..171S}, \citet{2018ApJ...856...16M}, and \citet{2019A&A...630A..79P} have shown that phenomena occurring in a plasma are more affected by ion-neutral collisions when the time and length scales of those phenomena are comparable to the corresponding collisional scales. For example, they have shown that the damping of waves and the plasma heating due to ion-neutral friction are more efficient when the wave frequencies are of the same order than the collision frequencies. On the other hand, the effect of these collisions is negligible when the oscillation frequency is much smaller than the frequency of the ion-neutral interaction. 
	
	For the present work, we need to compare the length-scale, $L_{0}$, and the numerical resolution, $\Delta x \equiv L_{0} / N_{0}$, with the collisional mean free path, $\lambda_{coll}$. This parameter can be estimated from Eq. (\ref{eq:nu_st}) in the following way:
	\begin{equation} \label{eq:mfpath}
		\lambda_{\rm{coll}} = \frac{V_{\rm{ref}}}{\nu_{\rm{st}}},
	\end{equation}
	where $V_{\rm{ref}}$ is a velocity representative of the dynamics of the system. Assuming the thermal speed given by Eq. (\ref{eq:vtherm}) as this reference value, the mean free path can be computed as
	\begin{equation} \label{eq:mfpath2}
		\lambda_{coll} \approx \frac{m_{\rm{s}}+m_{\rm{t}}}{n_{\rm{t}}m_{\rm{t}} \sigma_{\rm{st}}}.
	\end{equation} 
	With a cross-section of $\sigma_{\rm{in}} =1.16 \times 10^{-18} \  \rm{m^{2}}$ for the case of ion-neutral collisions \citep{2013PhPl...20f1202L}, we get a mean free path of the order of 10 to 100 m. This is two orders of magnitude smaller than the length-scale chosen for our simulations. Hence, the dynamics at the large scales should not be greatly affected by collisions. However, the opposite may be expected at the smaller scales, since the resolution would be comparable to the mean free path. For instance, if $N_{0} = 1000$, we have that $\Delta x = 10 \lesssim \lambda_{coll}$.
	
	Keeping the values of $\rho_{\rm{T1}}$, $T_{0}$, and $L_{0}$ mentioned in the previous paragraphs, we have performed several series of simulations to study the dependence of the results on various parameters and physical effects. In Sect. \ref{sec:sim_ref}, we present a detailed description of the dynamics of a small set of simulations that show the general evolution of the instability. In the remaining parts of this section we present studies on how the generation of magnetic field is affected by the presence of collisions and by the numerical resolution (Sect. \ref{sec:sim_eb}), and on the dependence on the ionisation degree (Sect. \ref{sec:sim_xc}).
	
\subsection{Reference simulations} \label{sec:sim_ref}
	We start by analysing in detail the case of a partially ionised plasma with an ionisation degree of $\chi_{\rm{c}} = 0.5$, that is, with the same values of density for the charged and the neutral fluid. The same flow speed is considered for the two fluids, with the heavier region moving at a speed $U_{0} = 5 \ \rm{km \ s^{-1}}$. This corresponds to Mach numbers of the charged and neutral fluids of
	\begin{equation} \label{eq:mach}
	    M_{\rm{c}} = \frac{U_{0}}{c_{\rm{S,c}}} \approx 0.3 \ \text{and} \ M_{\rm{n}} = \frac{U_{0}}{c_{\rm{S,n}}} \approx 0.43,
	\end{equation}
	respectively. Therefore, the flows are subsonic.
	
	The same perturbation is applied to the $z$-component of the velocity of both fluids, with an amplitude of $V_{z,0} = 10 \ \rm{m \ s^{-1}}$. However, this does not mean that charges and neutrals will have exactly the same dynamics. The reason is that, since they start with the same temperature, their pressures are different. This leads to some dissimilarities in their evolution, as we explain later.
	
	\begin{figure*}
		\centering
		\includegraphics[width=16.8cm]{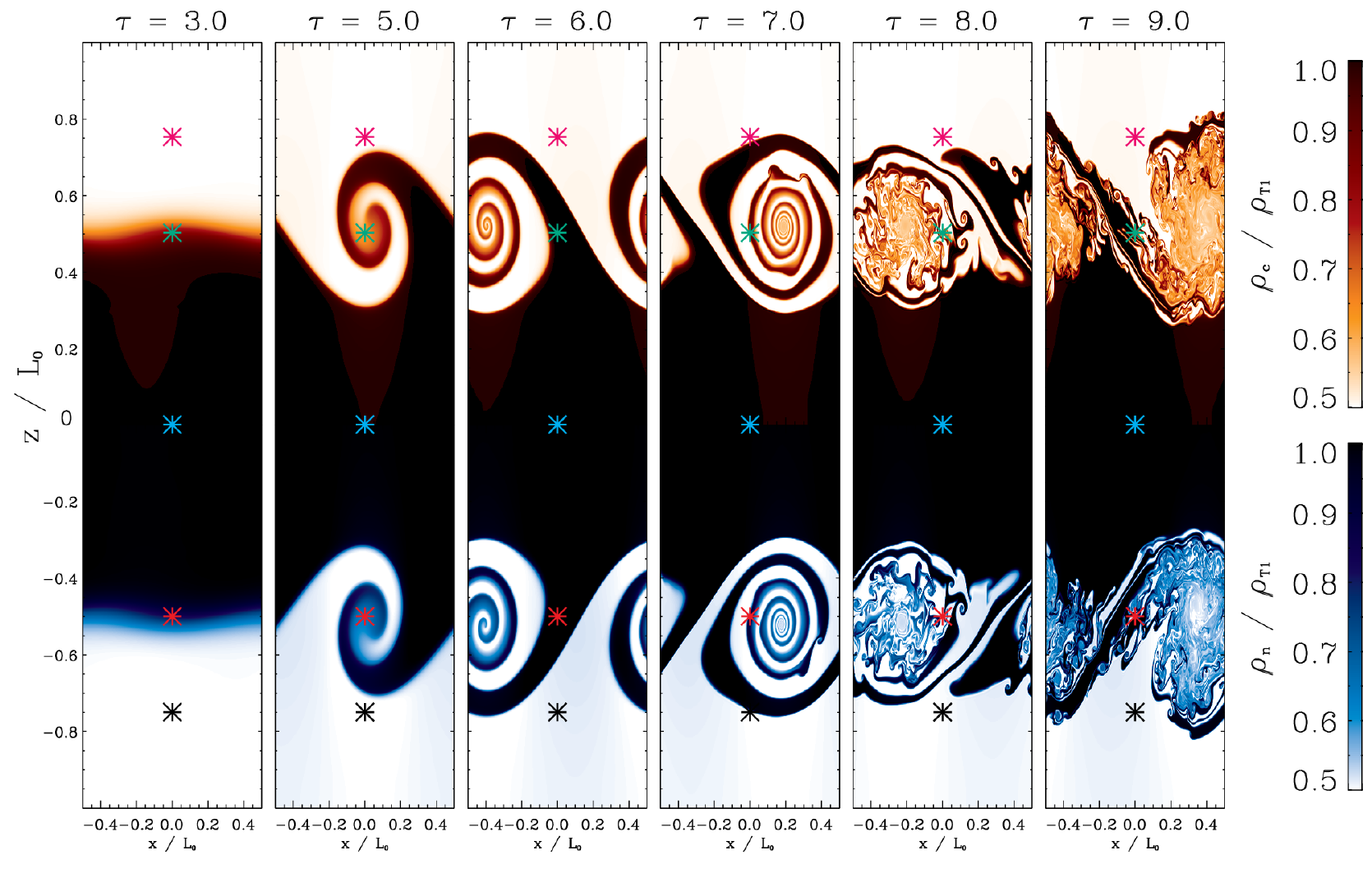}
		\caption{Density snapshots of a simulation with $\chi_{\rm{c}} = 0.5$ with no coupling between the charged (orange top half) and neutral (blue bottom half) fluids. Time is measured in the units of $\tau \equiv t / t_{0}$, where $t_{0}\equiv L_{0} / U_{0}$ is the domain crossing time. Asterisks represent key points whose temporal evolution is analysed in Sect. \ref{sec:points}. An animation of this figure is available online.}
		\label{fig:rhos}
	\end{figure*}
	
	\begin{figure*}
		\centering
		\includegraphics[width=16.8cm]{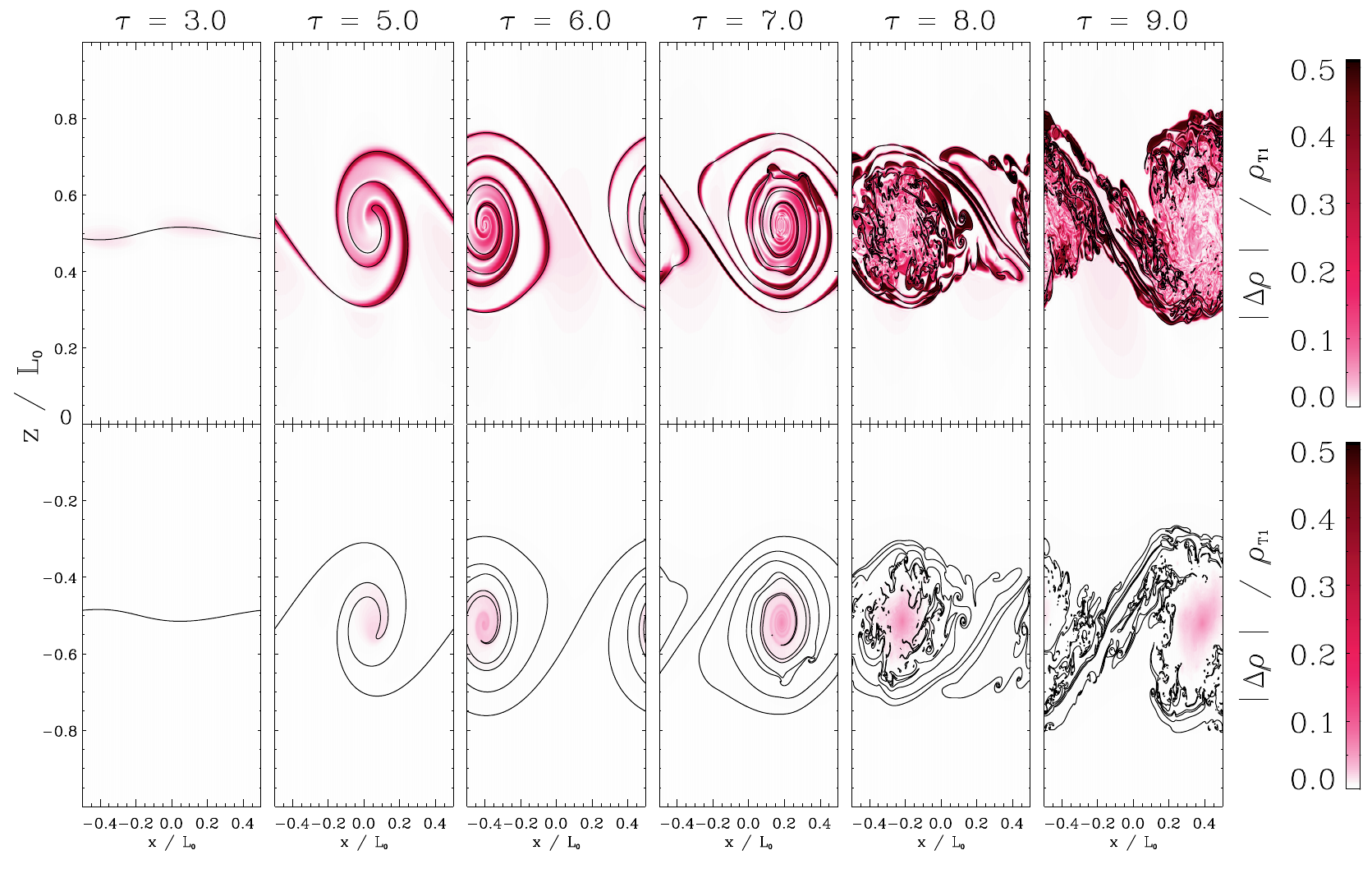}
		\caption{Differences in the evolution of density of charged and neutral fluids for the case of $\chi_{\rm{c}} = \chi_{\rm{n}} = 0.5$; $|\Delta \rho| \equiv |\rho_{\rm{c}} - \rho_{\rm{n}}|$. Top: $\alpha_{\rm{eff}} = 0$ (no collisions); bottom: $\alpha_{\rm{eff}} = \alpha$ (charge-exchange collisions not taken into account). The black lines represent density iso-contours.}
		\label{fig:rdiff}
	\end{figure*}

\subsubsection{Densities}
	Figure \ref{fig:rhos} shows the results of a simulation with $N_{0} = 1000$ points, in which the collisional interaction has not been taken into account ($\alpha_{\rm{eff}} = 0$). This means that there is no coupling between the charged and neutral fluids. Density snapshots at different times are presented, where $\tau \equiv t / t_{0}$ and $t_{0}$ is the domain crossing time, defined as $t_{0} \equiv L_{0} / U_{0}$. Due to the chosen initial setup, the evolution of the instability is symmetric with respect to $z = 0$. However, this figure only shows the results for the charged fluid in the upper half of the domain, and those for the neutral fluid in the lower half.  
	
	At first sight, it seems that the two-fluids have exactly the same dynamics: as time advances, the interfaces that separate regions of different densities are distorted because of the transfer of energy between the horizontal and vertical flows. During the initial steps, this distortion has the shape of the sinusoidal perturbation applied to the $z$-component of the velocities, whose amplitude increases exponentially, according to the linear analysis of \citet{1961hhs..book.....C}. At later stages, it turns into a large vortex \citep{1988FlDyR...3...93K}. Then, as energy cascades to smaller scales, secondary vortexes appear and, finally, a turbulent phase is reached \citep{2009JA014637}.
	
	However, if we examine Figure \ref{fig:rhos} with more attention, some small dissimilarities between the evolution of the charged and the neutral fluids can be found. For example, in the panels corresponding to $\tau = 6$ and $\tau = 7$, it can be checked that the main vortex is slightly more entangled in the case of the charged fluid. In addition, at later times, more small-scale structures such as secondary vortexes can be noticed on the upper half of the plot than in the lower half. 
	
	This different behaviour of the two fluids can be understood by taking into account the results of \citet{2012ApJ...749..163S}, who studied the effect of compressibility on the KHI. They have found that, for the case of small density ratios ($\rho_{\rm{T2}} / \rho_{\rm{T1}}$), compressibility reduces the growth rate of the instability in comparison with the incompressible limit. This limit is obtained by assuming that the sound speed tends to infinity.
	
	Since in our simulations $c_{\rm{S,c}} > c_{\rm{S,n}}$, compressibility should have a stronger effect on the neutral than in the charged fluid, and the growth rates of the instability should be slightly larger for the latter. This difference on the growth rates explains why the charged fluid is slightly faster in reaching the non-linear and turbulent stages of the KHI.

	The existence of the above mentioned differences is more clearly revealed by defining the variable $|\Delta \rho | \equiv |\rho_{\rm{c}} - \rho_{\rm{n}}|$. We show in the upper half of Fig. \ref{fig:rdiff} snapshots of this variable that correspond to the snapshots shown in Fig. \ref{fig:rhos}. During the linear phase, the differences are very small, as it can be checked by the fact that the panel for $\tau = 3$ is almost empty (small patches of light grey can be found around $z / L_{0} = 0.5$). Later, the differences become more evident: they are concentrated at the edges of the main vortex. But as time advances and turbulence comes into play, they appear in a larger region of the domain.
	
	The bottom half of Fig. \ref{fig:rdiff} shows the results for a simulation that includes elastic collisions between the charged and neutral fluids ($\alpha_{\rm{eff}} = \alpha$). Compared to the previous case, it can be seen that the dynamics of both fluids are much more similar. Nevertheless, there are still some differences in the evolution of densities that reveal that the coupling between the two fluids is not perfect. These differences have a relative magnitude of less than 0.1, which is notably smaller than the ones for the case without collisions, and are mainly found in the small-scale structures, in the cores of the vortexes. On the other hand, the two fluids are strongly coupled in the large scales. These results are in good agreement with the findings of \citet{2019PhPl...26h2902H}.
	
	\begin{figure*}
		\centering
		\includegraphics[width=17cm]{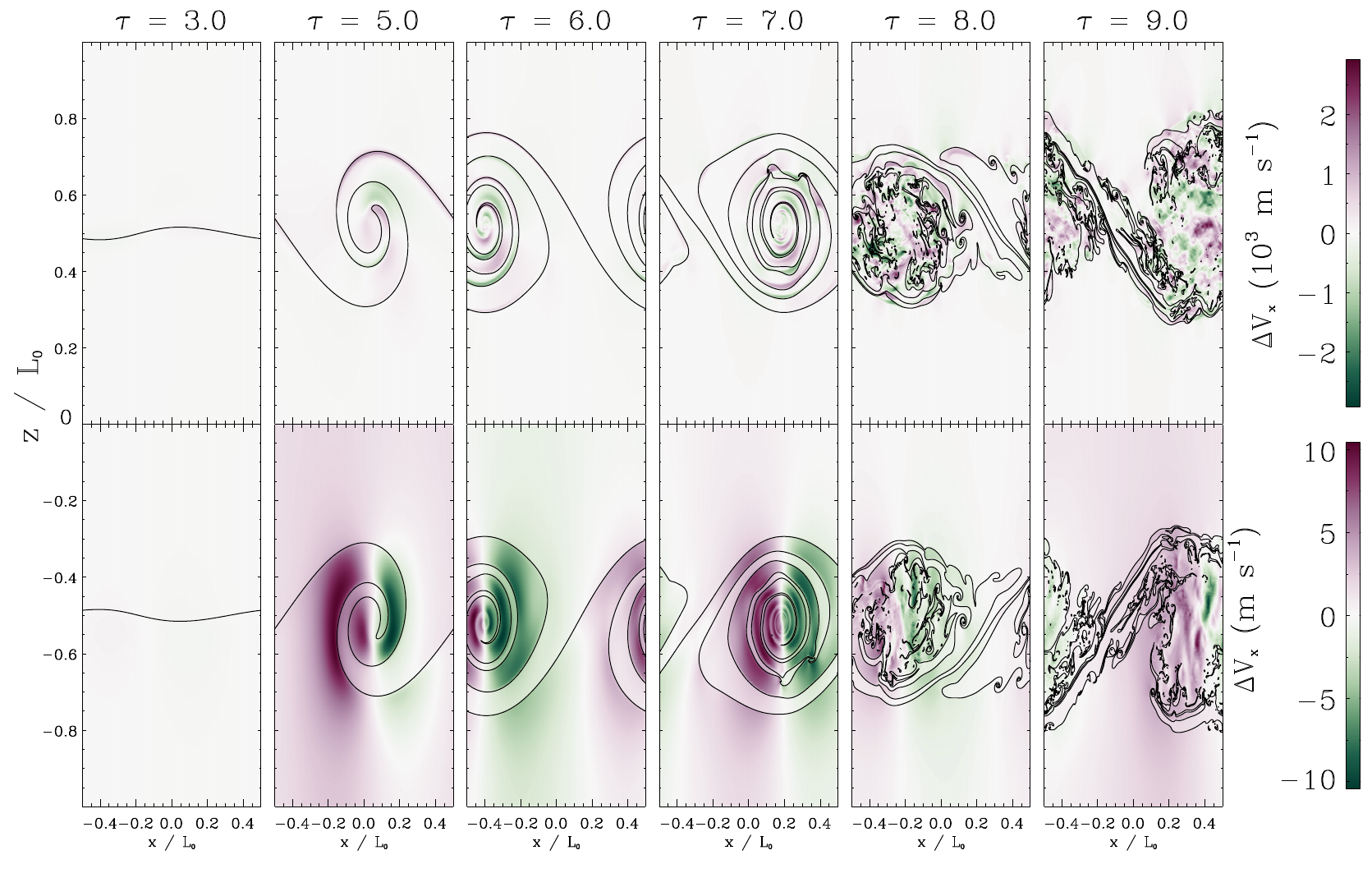}
		\caption{Velocity drifts, $\Delta V_{x} \equiv V_{x\rm{,c}} - V_{x\rm{,n}}$, for the same simulations as in Fig. \ref{fig:rdiff}. The black lines represent density iso-contours. We note that the units of the top colourbar are $\rm{km \ s^{-1}}$ and the units of the bottom colourbar are $\rm{m \ s^{-1}}$: the presence of the collisional coupling reduces the magnitude of the velocity drifts.}
		\label{fig:vdiff}
	\end{figure*}

\subsubsection{Velocity drifts}
	Another useful variable to quantify the effect of collisions on the dynamics of the plasma is the velocity drift, $v_{\rm{D}}$. When this variable tends to zero, the two fluids are perfectly coupled and they have the same dynamics, which can be precisely described by single-fluid models. If $v_{\rm{D}} \ne 0$ (and collisions are considered), there is a transfer of momentum and energy between the two species, as shown by Eqs. (\ref{eq:Rcoll}), (\ref{eq:qcn}) and (\ref{eq:qnc}).
	
	In Fig. \ref{fig:vdiff} we present snapshots of the $x$-component of the velocity drift, $\Delta V_{x} \equiv V_{x\rm{,c}} - V_{x\rm{,n}}$. The top panels of Fig. \ref{fig:vdiff} show the results for the case without collisions, while the bottom panels correspond to the simulation that includes elastic collisions. As expected, the behaviour shown by the top panels is similar to the one presented in Fig. \ref{fig:rdiff} for the density: during the first stages of the simulation, the main differences in velocity appear around the vortex edges but later they extend to the whole turbulent mixing region. The maximum velocity drifts can be found in the later stages of the evolution and in the smaller-scales, with values of the order of $2.5 \ \rm{km \ s^{-1}}$. During the whole simulation there are zones in which the charged fluid is moving faster than the neutral one and other in which it is slower. But there is no clear trend on the spatial distribution of these areas.
	
	On the contrary, a trend can be distinguished in the bottom panels of Fig. \ref{fig:vdiff}, which represent the case with charged-neutral collisions. Looking to the panel for $\tau = 5$, it can be seen that in the leading part of the main vortex the velocity drifts are negative, which means that the neutral fluid has a larger horizontal speed than the charged fluid. The opposite behaviour is found in the trailing part. This trend is maintained during the remaining of the simulation, even at the turbulent stage. Although we do not include here the corresponding plot, we find a different trend in the $z$-component of the velocity drifts: in this case, the neutral fluid is moving faster in the upper part of the main vortex and slower in the lower part.
	
	By comparing the bottom panels of Figs. \ref{fig:rhos} and \ref{fig:vdiff}, it can be checked that the largest velocity drifts are found in the regions with the lowest densities. This can be explained by the expression of the momentum transfer. According to Eq. (\ref{eq:Rcoll}), the momentum transfer is proportional to the density of both fluids. Therefore, a lower collisional friction should be expected in regions with lower densities, which leads to a weaker coupling and larger values of the velocity drifts. As shown by the bottom colourbar of Fig. \ref{fig:vdiff}, these maximum values of $v_{\rm{D}}$ are of the order of $10 \ \rm{m \ s^{-1}}$, two orders of magnitude smaller than in the case without collisions. They are also much smaller than the thermal speeds of the two fluids, in agreement with the range of applicability of the momentum and energy transfer expressions given by Eqs. (\ref{eq:Rcoll}), (\ref{eq:qcn}) and (\ref{eq:qnc}) \citep{1977RvGSP..15..429S,1986MNRAS.220..133D}.
	
\subsubsection{Magnetic field and temperature at key points} \label{sec:points}
    Now we pay attention to the temporal evolution of some selected points of the numerical domain. The chosen points are marked as asterisks in Figure \ref{fig:rhos}. We denote as $P1$ the point at the internal denser region of the plasma, represented as a light blue asterisk. Points $P2$ and $P3$, represented by red and green symbols, respectively, are initially located at the transition layers between the internal and external layers of the plasma. Finally, points $P4$ (black) and $P5$ (pink) are located at the external lighter regions.

	\begin{figure}
		\centering
		\resizebox{\hsize}{!}{\includegraphics[]{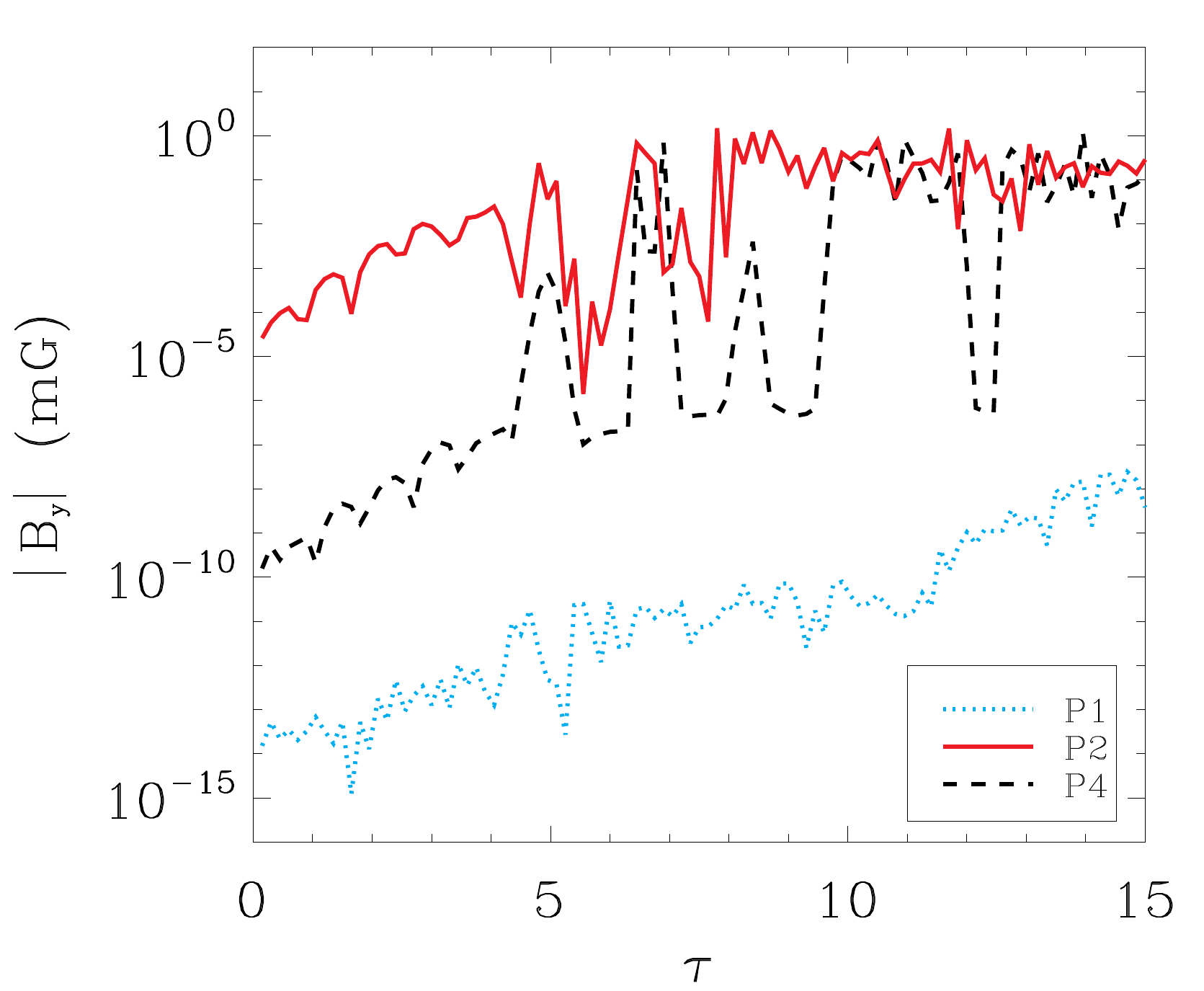}}
		\caption{Absolute value of the magnetic field, $|B_{y}|$, at points $P1$ (blue dotted line), $P2$ (red solid line) and $P4$ (black dashed line) as a function of the normalised time, $\tau$.}
		\label{fig:by_points}
	\end{figure}
	
	\begin{figure*}
		\centering
		\includegraphics[width=0.495\hsize]{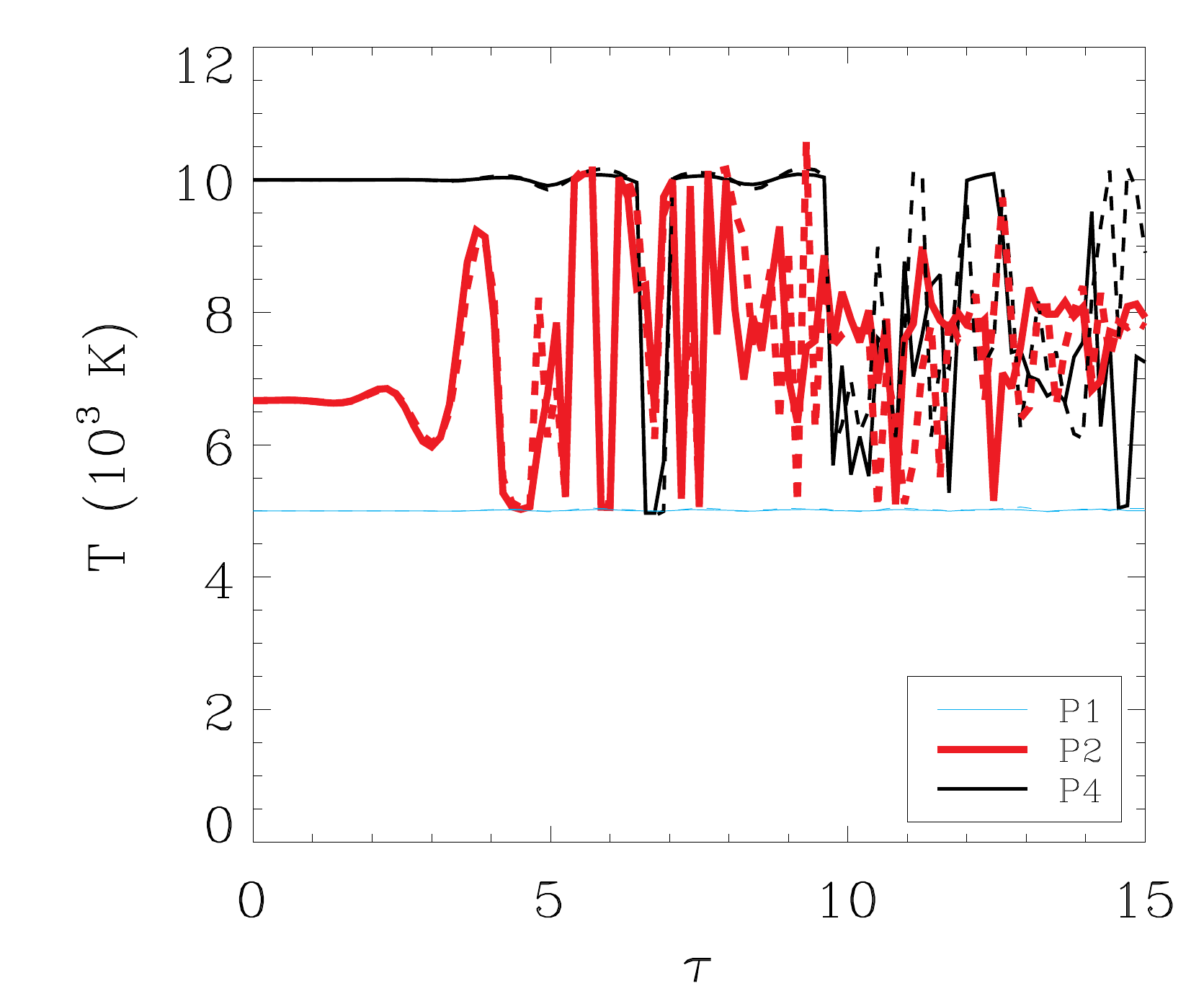}
		\includegraphics[width=0.495\hsize]{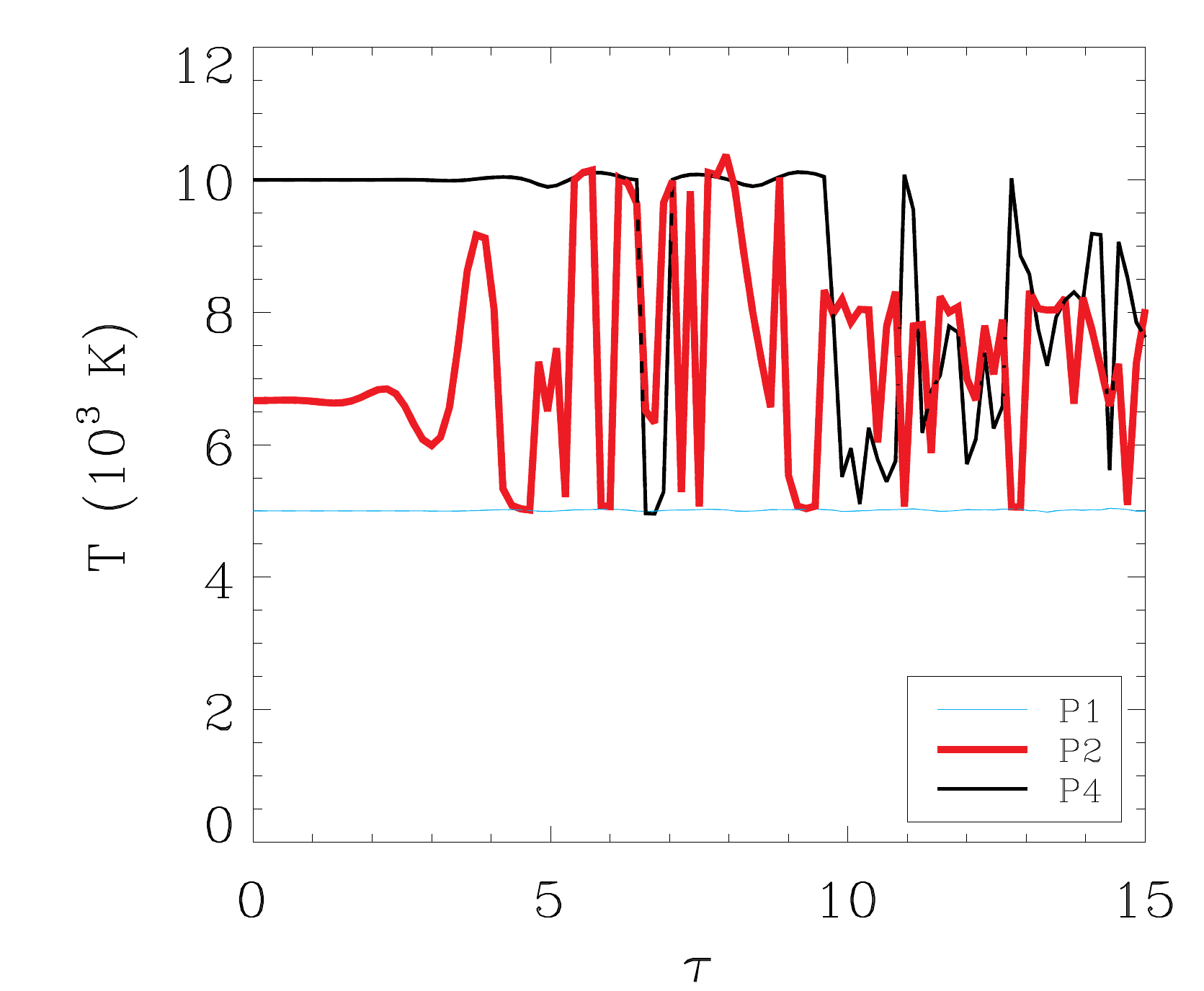}
		\caption{Temperature as a function of time at points $P1$ (blue thin line), $P2$ (red thick line) and $P4$ (black line). Solid lines represent the temperature of the charged fluid while dashed lines represent the temperature of the neutral fluid. Left and right panels correspond to the cases with $\alpha_{\rm{eff}} = 0$ and $\alpha_{\rm{eff}} = \alpha$.}
		\label{fig:tt_points}
	\end{figure*}
	
	Figure \ref{fig:by_points} shows the evolution of the absolute value of the magnetic field at points $P1$, $P2$ and $P4$. Qualitatively, the magnetic field at $P2$ and $P4$ follows a similar evolution, which can be clearly separated in two stages: exponential growth and saturation (with even a slight decrease). However, during the first stage, at $P2$ the magnetic field is several orders of magnitude larger than at the other point. This behaviour is consistent with what is expected from the Biermann battery mechanism: at the transition layers there are larger gradients of pressure and density than in the other regions of the domain. Later, during the saturation stage, the magnetic field at $P4$ reaches the same order of magnitude than at $P2$. This is a consequence of the non-linear evolution of the KHI, which creates a mixing layer that, in this case, contains the two selected points. Then, at the centre of the domain ($P1$), the magnetic field is negligible in comparison with the other points, although it has a different evolution: for the particular case of this simulation, it keeps growing with time without reaching a saturation phase. We have checked that in other configurations, with larger values of numerical diffusivities, the saturation stage is indeed reached by all the selected points.
	
	The magnetic field at points $P3$ and $P5$ has not been represented on Fig. \ref{fig:by_points} because we have checked that it has the same absolute value as for $P2$ and $P4$, respectively, although the actual value has the opposite sign.
	
	The evolution of the temperature at the points $P1$, $P2$ and $P4$ is shown in Fig. \ref{fig:tt_points} (in the same way as for Fig. \ref{fig:by_points}, the results for $P3$ and $P5$ are not shown here because they are equivalent to those of $P2$ and $P4$, respectively). Initially, the plasma at $P1$ (the denser region) has a temperature of $5 \times 10^{3} \ \rm{K}$, while the temperatures at $P2$ (transition layer) and $P4$ (lighter region) are $\sim 6600 \ \rm{K}$ and $10^{4} \ \rm{K}$, respectively. During the whole simulation the temperature at $P1$ is almost the same, with very small variations. That is not the case for $P2$ and $P4$. As the instability develops and the plasma is carried by the flow, those static points are crossed by plasma that corresponds to regions with different densities and temperatures, which explains the large fluctuations of temperature at $P2$. The temperature at $P4$ remains almost constant for a larger time than for $P2$ due to the time needed for the development of the mixing layer. Finally, once this mixing layer is created, both points have a similar temperature, which is higher than the initial temperature at $P2$ but lower than the initial temperature of the external region.
	
	In addition, the left panel of Fig. \ref{fig:tt_points}, which represents a case with no ion-neutral collisions, shows that appreciable differences in the temperatures of the two fluids are present during the turbulent stage. In contrast, when collisions are taken into account these large temperature differences are removed, as shown by the right panel, in which the solid and dashed lines overlap.
	
\subsection{Magnetic field: growth rate and effect of resolution} \label{sec:sim_eb}
	In this section we focus on the evolution of the magnetic field by analysing its average value over the whole domain instead of its evolution at some given points. The average of the magnetic field is computed as follows:
	\begin{equation} \label{eq:eb_av}
		\langle|\bm{B}|\rangle = \sum_{i=1}^{N_{0}}\sum_{j=1}^{2N_{0}}\frac{\sqrt{B_{x}(i,j)^{2}+B_{y}(i,j)^{2}+B_{z}(i,j)^{2}}}{2N_{0}^{2}}.
	\end{equation}
	
	\begin{figure*}
		\centering
		\includegraphics[width=0.495\hsize]{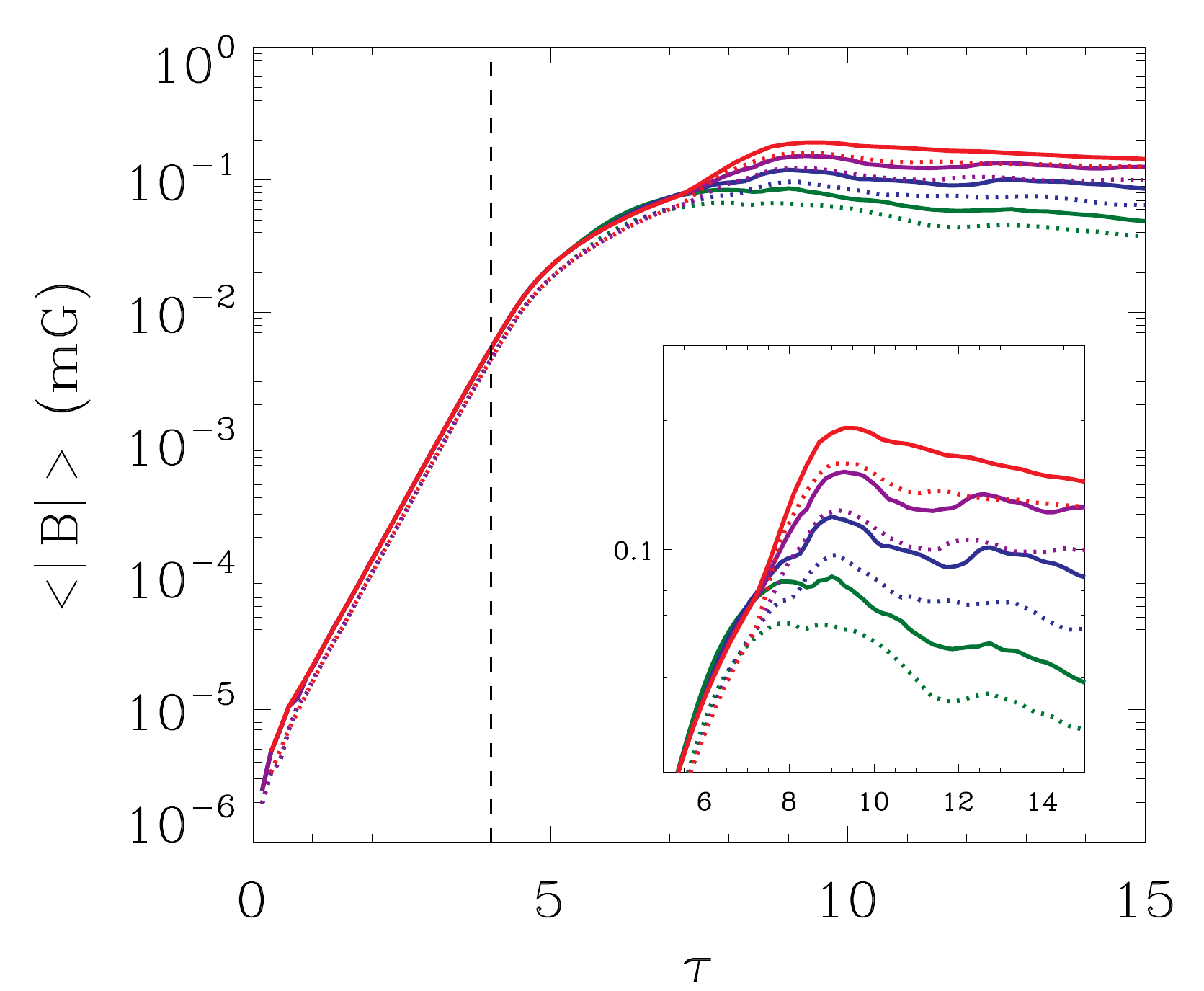}
		\includegraphics[width=0.495\hsize]{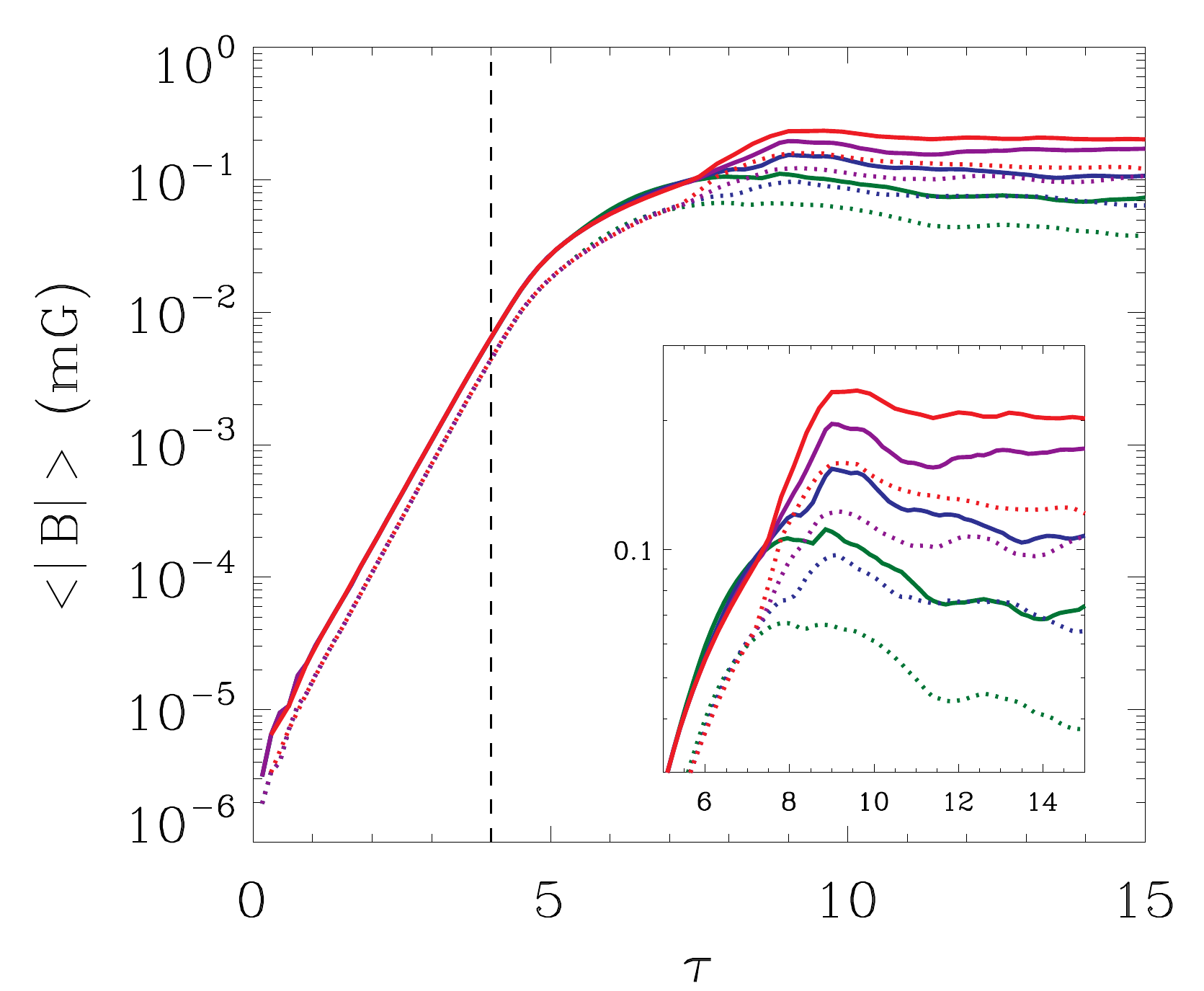}
		\caption{Averaged magnetic field as a function of time and resolution for the cases with $\chi_{\rm{c}} = 0.5$ (left) and $\chi_{\rm{c}} = 0.1$ (right). Solid lines represent simulations in which the ion-neutral collisional interaction is taken into account (without charge-exchange). Dotted lines represent simulations without collisional coupling. Green, blue, purple and red colours correspond to simulations with $N_{0} = 250, 500, 1000, 2000 $, respectively. The vertical dashed lines approximately mark the separation between the linear and non-linear phases of the instability. The inset in each panel shows a zoom of the interval $\tau \in [5,15]$.} 
		\label{fig:eb_resol}
	\end{figure*}
	
	Figure \ref{fig:eb_resol} shows that the spatially averaged magnetic field initially follows a linear phase in which its magnitude increases exponentially with time: from $\tau = 0$ to $\tau = 4$, $\langle |\bm{B}| \rangle$ rises up by $4$ orders of magnitude. Then, it keeps increasing but at a lower rate until it reaches a maximum around $\tau = 9$ and starts decreasing due to the effect of the physical and numerical dissipation mechanisms. We can compute the normalised growth rate of the magnetic field during the linear phase using the following formula:
	\begin{equation}
		\Gamma_{B} = \frac{\ln \langle|\bm{B}(\tau_{2})| \rangle -\ln \langle |\bm{B} (\tau_{1})|\rangle}{\tau_{2}-\tau_{1}}.
	\end{equation}
	This produces a result of $\Gamma_{B} \approx 2.02$.

	According to \citet{1961hhs..book.....C}, the linear growth rate of the KHI (which is expressed in units of $s^{-1}$) is given by 
	\begin{equation}
		\gamma_{\rm{th}} = \frac{2 \pi}{\lambda}\Delta U \frac{\sqrt{\rho_{\rm{T1}}\rho_{\rm{T2}}}}{\rho_{\rm{T1}}+\rho_{\rm{T2}}},
	\end{equation}
	which states that perturbations with larger wavelengths, $\lambda$, grow slower. For the case of our simulations, the growth rate of the initial perturbation (with $\lambda = L_{0}$) can be expressed in normalised units as
	\begin{equation}
		\Gamma_{\rm{th}} = \gamma_{\rm{th}} t_{0} =\frac{\gamma_{\rm{th}} L_{0}}{U_{0}} = \frac{\gamma_{\rm{th}}\lambda}{\Delta U}= 2 \pi \frac{\sqrt{\rho_{\rm{T1}} \rho_{\rm{T2}}}}{\rho_{\rm{T1}} + \rho_{\rm{T2}}} \approx 2.96,
	\end{equation}
	which is larger than the one found in the simulations. This discrepancy can be explained due to the fact that the analytical linear growth rate, $\Gamma_{\rm{th}}$, was computed assuming a sharp interface between the two media, while in our simulations there is a smooth variation of density and velocity. The smooth transition reduces the growth rate of the instability, as shown by, for instance, \citet{1982JGR....87.7431M} \citet{2004hyst.book.....D} or \citet{2019MNRAS.485..908B}. Using a model in which the velocity varies linearly between the inner and outer layers in a scale of $2 w_{L}$, \citet{2004hyst.book.....D} derived an analytic expression for the growth rate for the perturbations in velocity or magnetic field. With the notation employed in the present work, the mentioned expression is given by
	\begin{equation}
		\gamma_{\rm{DR}} = \operatorname{Im}\left[\frac{\Delta U}{4 w_{\rm{L}}}\sqrt{\left(\frac{4\pi}{\lambda}w_{\rm{L}} - 1\right)^{2}-\exp\left(-\frac{8\pi}{\lambda}w_{\rm{L}}\right)}\right].
	\end{equation} 
	Inserting the parameters of our simulations in the formula above, we get a result of $\gamma_{\rm{DR}} \approx 1.08$. Then, the growth rate for the magnetic field, given in normalised units, would be $\Gamma_{\rm{DR}} = \gamma_{\rm{DR}} t_{0} \approx 2.16$, which is closer to the value computed from the numerical simulation.
	
	The different colours in Fig. \ref{fig:eb_resol} represent simulations with different values of the resolution parameter $N_{0}$: $250$ (green), $500$ (blue), $1000$ (purple), and $2000$ (red).  It can be seen that the results of the four resolutions coincide during the linear phase of the instability. However, they start to diverge once the non-linear stage is reached. As a larger number of points is used and, hence, smaller scales can be better resolved, the generation of magnetic energy is larger and the peak value of magnetic field is reached at a later time. This is mainly due to the effect of the numerical dissipation, which depends on the grid size of the numerical domain, see Eq. (\ref{eq:nu_l}): improving the resolution of the simulation decreases the numerical dissipation. Furthermore, it takes longer for the instability to cascade its energy to the dissipation scales in a simulation with a smaller grid size. 
	
	In addition, the dotted lines of Fig. \ref{fig:eb_resol} correspond to simulations in which the charged-neutral collisions were neglected, and the solid lines correspond to the case with only elastic collisions (no charge-exchange). According to these results, larger magnetic fields are generated when the charged-neutral collisions are taken into account. This effect is studied in more detail in the next section.

\subsection{Effect of collisions and dependence on ionisation degree} \label{sec:sim_xc}
	\begin{figure*}
		\centering
		\includegraphics[width=0.329\hsize]{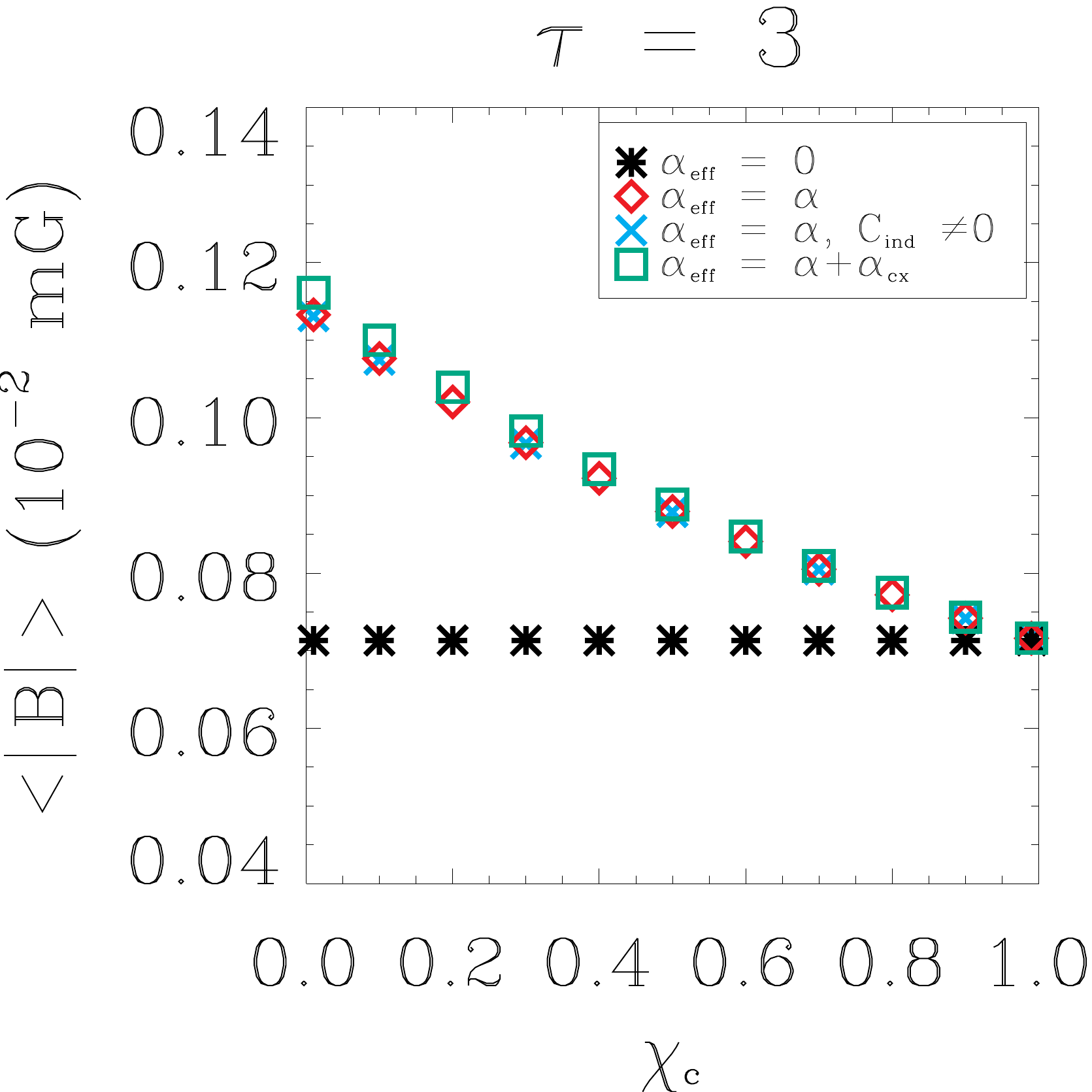}
		\includegraphics[width=0.329\hsize]{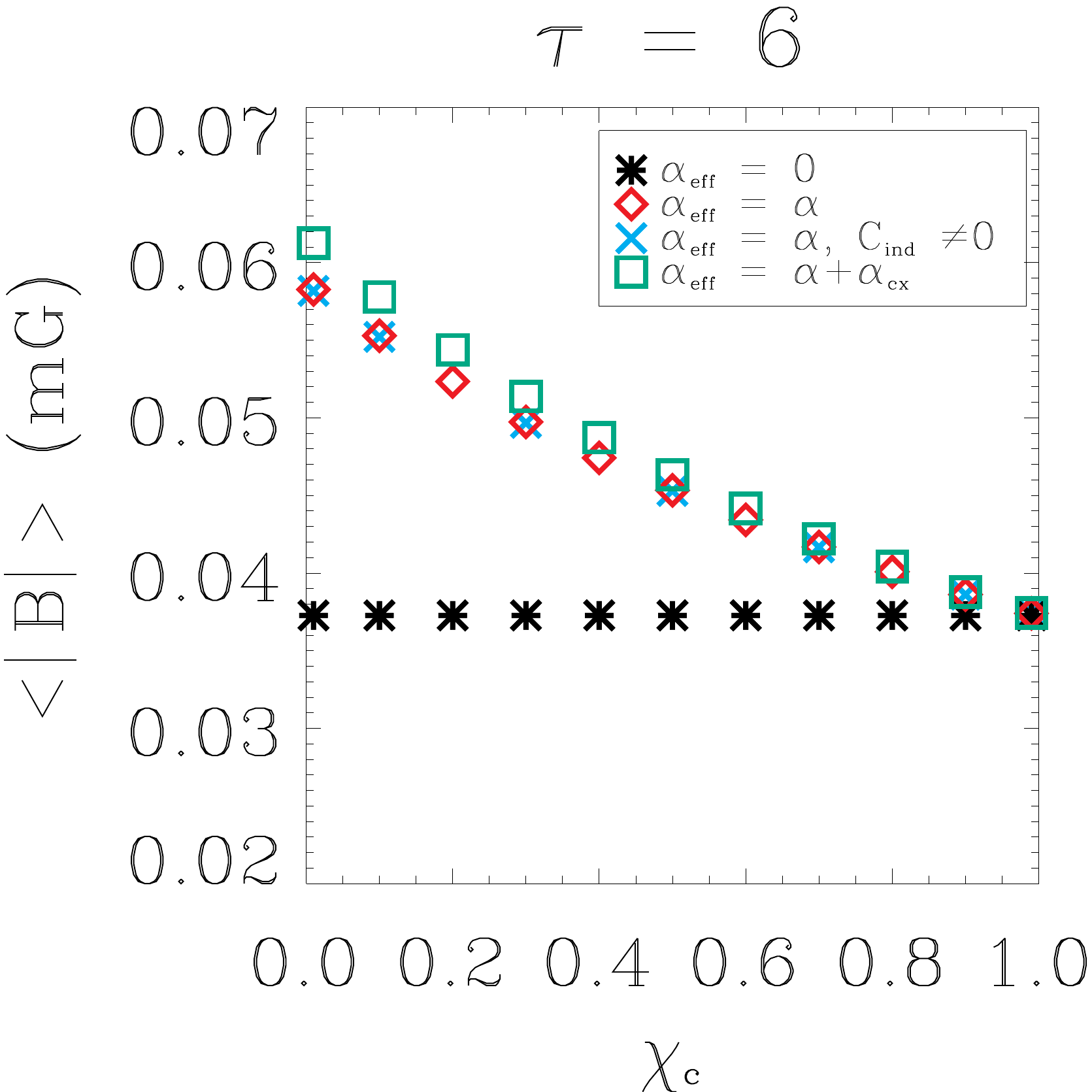}
		\includegraphics[width=0.329\hsize]{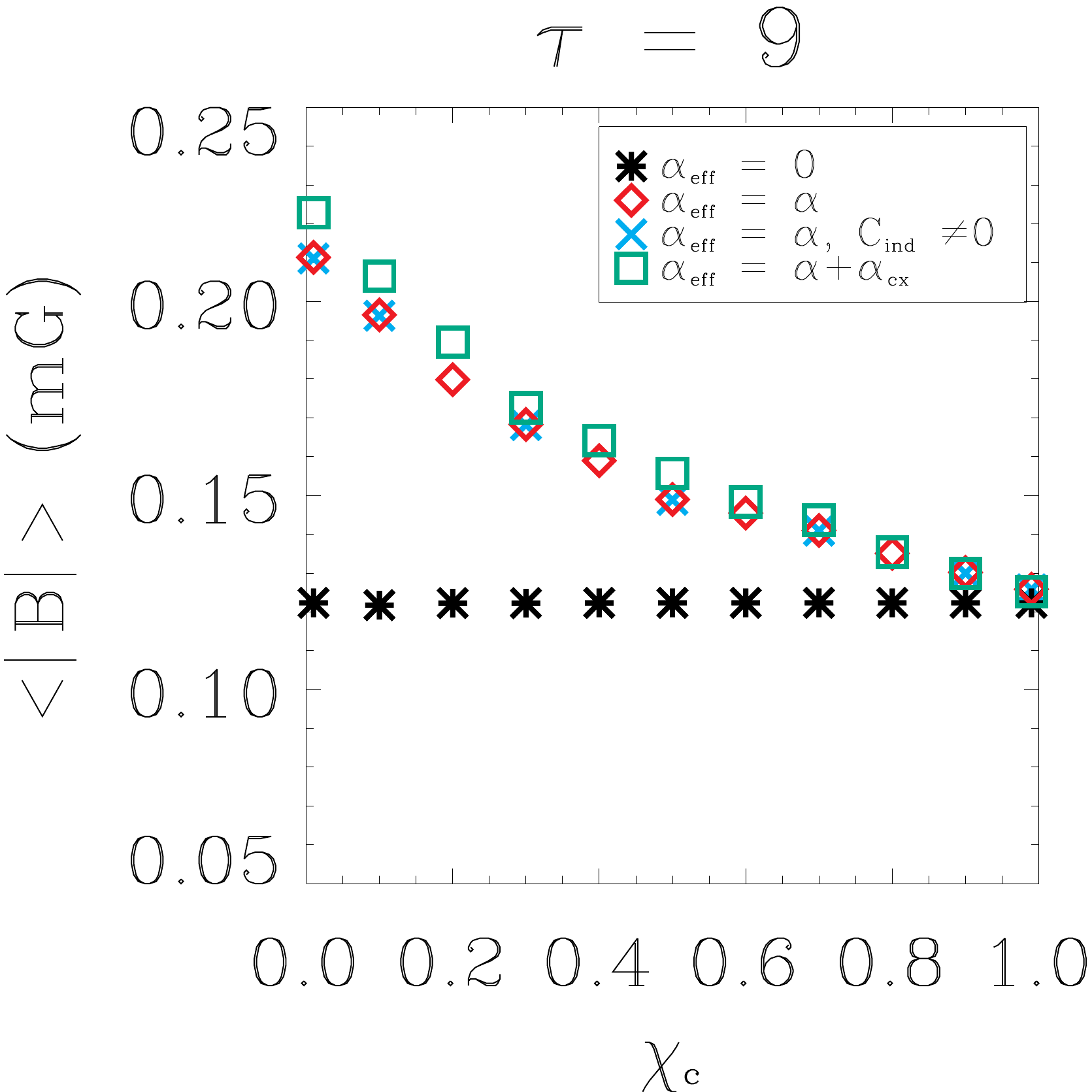}
		\caption{Spatially averaged magnetic field as a function of ionisation degree at three different times of the simulations. Black asterisks show the results without collisions. Red diamonds correspond to simulations in which ion-neutral collisions are considered. Blue crosses represent the results when the collisional term of the induction equation is also taken into account and the green squares represent the simulations that also included the process of charge-exchange.}
		\label{fig:bfield_xc}
	\end{figure*}

	The comparison of the left and right panels of Fig. \ref{fig:eb_resol}, which correspond to plasmas with ionisation degrees of $\chi_{\rm{c}} = 0.5$ and $\chi_{\rm{c}} = 0.1$, respectively, shows that the effect of collisions on the generation of magnetic field varies with the ionisation degree. To investigate such a dependence, we have performed four different series of simulations. The details of those simulations are summarised in Table \ref{table:sims}. Within each series we have varied the ionisation degree from a minimum value of $\chi_{\rm{c}} = 0.01$ to a maximum of $\chi_{\rm{c}} = 0.99$. The rest of parameters, such as the total density, temperatures and velocities, are the same as in the previous section.

    \begin{table} [h]
		\caption{Series of simulations}
		\label{table:sims}
		\centering
		\begin{tabular}{c c  c c c}
			\hline\hline
			Tag & $\alpha_{\rm{eff}}$ & $\bm{C}_{\rm{Ind}}$ & Symbol \\
			\hline
			I & 0 & No & $\textcolor{black}{\ast}$ \\
			II & $\alpha$ & No & $\textcolor{red}{\diamond}$ \\
			III & $\alpha $ & Yes & $\textcolor{cyan}{\times}$ \\
			IV & $\alpha + \alpha_{\rm{cx}}$ & Yes & $\textcolor{green}{\square}$\\
			\hline
		\end{tabular}
		\tablefoot{The third column indicates whether the collisional term from the induction equation has been taken into account or not. The last column indicates the symbol that represents each series in Fig. \ref{fig:bfield_xc}.}
	\end{table}
	
	Figure \ref{fig:bfield_xc} displays three panels that correspond to different stages of the evolution of the instability. At $\tau = 3$ (left panel) the instability is still in the linear phase (as shown in Fig. \ref{fig:eb_resol}). At time $\tau = 6$ (middle panel), the magnetic field continues to grow but it does not follow a linear dependence with time. The right panel represents the time $\tau = 9$, in which the plasma is already in a turbulent stage (see the rightmost panel of Fig. \ref{fig:rhos}), and corresponds approximately to the moment when the magnetic energy reaches its maximum value (saturation).

	The magnitudes of the averaged magnetic field at each stage are quite different (note the scales in each panel) but the general dependence on the ionisation degree is the same. When no collisions between the charged and the neutral fluids are considered (series I), the resulting magnetic field does not vary with the ionisation degree. The reason is that, in the uncoupled scenario, the generation of magnetic energy is related only to the evolution of the KHI in the charged fluid. Apart from the value of the shear flow velocity and the wavelength of the initial perturbation, which are the same for every simulation in the series, the instability depends on the density ratio between the heavier and lighter media and the sound speed. In our simulations, neither of these parameters are functions of the ionisation degree. So, the same dynamics should be expected throughout this first series.
	
	When the charged-neutral interaction comes into play, Fig. \ref{fig:bfield_xc} shows a remarkable dependence on the ionisation degree. A larger magnetic field is obtained as the ionisation degree is decreased, that is, as the plasma becomes more weakly ionised. And the variation with respect to the uncoupled case is clearly not negligible. For a partially ionised plasma with $\chi_{\rm{c}} = 0.5$, the variation is of about $27 \ \%$, while for a weakly ionised plasma, with $\chi_{\rm{c}} = 0.01$, it reaches up to a $\sim 80 \ \%$.
	
	Comparing the results of the three series that included collisional terms, it can be checked that the effect of the collisional term of the induction equation is negligible: the results for the case that includes that term (series III) almost overlap with the results corresponding to the case with only elastic collisions (series II). On the other hand, the inclusion of charge-exchange collisions (series IV) slightly increases the averaged magnetic field, specially in the weakly ionised regime.
	
	The physical reasons behind the results commented in the paragraphs above are explained in Sect. \ref{sec:disc}.

\section{Discussion. Analysis of the equations} \label{sec:disc}
	To better understand the results of the simulations described in the previous section, it is necessary to analyse in more detail the equations included in the two-fluid model. In the present section we pay special attention to the Biermann battery term of the induction equation and the collisional terms that appear in the momentum and the energy equations.
	
\subsection{Induction equation} \label{sec:ind}
    First, we analyse the case without the collisional interaction between charged and neutral particles. Since our simulations start without an initial seed, the very first steps of the evolution of the magnetic field are described by Eq. (\ref{eq:indu_batt}), which can also be written as
  	\begin{equation} \label{eq:batt_c}
  		\frac{\partial \bm{B}}{\partial t}=-\frac{m_{\rm{i}}\nabla \rho_{\rm{c}} \times \nabla P_{\rm{c}}}{2e \rho_{\rm{c}}^{2}} = -\frac{k_{\rm{B}}\nabla \rho_{\rm{T}} \times \nabla T_{\rm{c}}}{e \rho_{\rm{T}}}.
 	\end{equation}
 	Here we have used the facts that $\rho_{\rm{c}} = \chi_{\rm{c}} \rho_{\rm{T}}$, $P_{\rm{c}} = 2 \rho_{\rm{c}} / m_{\rm{i}} k_{\rm{B}} T_{\rm{c}}$ and $\chi_{\rm{c}}$ is uniform. With this formula we can explain the behaviour presented in Fig. \ref{fig:bfield_xc} for the series without collisions (black asterisks). Under the conditions of that series, the evolution of the magnetic field is independent from the ionisation degree of the plasma.
  	
  	In addition, if we take into account that in our model there are no variations along the $y$-direction, that is, $\partial y = 0$, from Eq. (\ref{eq:batt_c}) we obtain that the battery term only generates magnetic field in the $y$-direction, so $B_{x} = B_{z} = 0$. The evolution of the $y$-component of the magnetic field is given by
  	\begin{equation}
  		\frac{\partial B_{y}}{\partial t} = \frac{m_{\rm{i}}}{2e \rho_{\rm{c}}^{2}} 
  	   		\big[\left(\partial_{x} \rho_{\rm{c}}\right) \left(\partial_{z} P_{\rm{c}}\right)-\left(\partial_{z}\rho_{\rm{c}}\right) \left(\partial_{x} P_{\rm{c}}\right)\big]
  	\end{equation}
  	where $\partial_{x}$ and $\partial_{z}$ represent the derivatives with to $x$ and to $z$, respectively.
  	
  	Once a seed magnetic field appears due to the action of the Biermann battery term, the advection term of the induction equation becomes different from zero. Taking also into account that $V_{y,\rm{c}}=0$, the evolution of the magnetic field due to that term is given by
  	\begin{equation}
  		\frac{\partial B_{y}}{\partial t} =
  		  			-\partial_{x} \left(V_{x\rm{,c}}B_{y}\right)-\partial_{z} \left(V_{z\rm{,c}}B_{y}\right).
  	\end{equation}
  	
 	The fact that no magnetic field is present in the $x$ and $z$ directions is important for the evolution of the instability. It is known that a magnetic field that is oriented along the direction of the shear flow velocity has a stabilizing effect \citep{1961hhs..book.....C}. In this case, the generated magnetic field is perpendicular to that flow, so it has no effect on the KHI.
 	
\subsubsection{Vorticity} \label{sec:eq_vort}
  	The vorticity of any species ``s'' is defined as the curl of its velocity, that is, $\bm{\omega}_{\rm{s}} \equiv \nabla \times \bm{V}_{\rm{s}}$. By operating the continuity and momentum equations, Eqs. (\ref{eq:rho}--\ref{eq:momn}), we obtain  equations of evolution for the vorticities of both fluids:
 
  	\begin{eqnarray} \label{eq:vort_c}
  		\lefteqn{\frac{\partial \bm{\omega}_{\rm{c}}}{\partial t} = \nabla \times \left(\bm{V}_{\rm{c}} \times \bm{\omega}_{\rm{c}} \right) + \frac{\nabla \rho_{\rm{c}} \times \nabla P_{\rm{c}}}{\rho_{\rm{c}}^{2}}} \nonumber \\ 
  		& & +\frac1{\mu_{0}} \nabla \times \left[\frac{(\nabla \times \bm{B}) \times \bm{B}}{\rho_{\rm{c}}}\right] + \nabla \times \left[\nu_{\rm{cn}} \left(\bm{V}_{\rm{n}}-\bm{V}_{\rm{c}} \right) \right] 
  	\end{eqnarray}
  	and
  	\begin{eqnarray} \label{eq:vort_n}
  		\lefteqn{\frac{\partial \bm{\omega}_{\rm{n}}}{\partial t} = \nabla \times \left(\bm{V}_{\rm{n}} \times \bm{\omega}_{\rm{n}} \right) + \frac{\nabla \rho_{\rm{n}} \times \nabla P_{\rm{n}}}{\rho_{\rm{n}}^{2}}} \nonumber \\
  		& & + \nabla \times \left[\nu_{\rm{nc}} \left(\bm{V}_{\rm{c}}-\bm{V}_{\rm{n}}\right) \right].
  	\end{eqnarray}
  	
  	It can be seen that the vorticity equation for the charged fluid has a great resemblance with the induction equation. So, in the same way, we rewrite it as
  	\begin{equation}
  		\frac{\partial \bm{\omega}_{\rm{c}}}{\partial t} = \bm{A}_{\omega \rm{c}} + \bm{B}_{\omega \rm{c}} + \bm{L}_{\omega \rm{c}} + \bm{C}_{\omega \rm{c}},
  	\end{equation}
  	where $\bm{A}_{\omega \rm{c}}$ is the advective term, $\bm{B}_{\omega \rm{c}}$ is called the baroclinic term, $\bm{L}_{\omega \rm{c}}$ is related to the Lorentz force, and $\bm{C}_{\omega \rm{c}}$ refers to the collisional interaction between the charged and neutral fluids. The same procedure can be applied to the neutral vorticity equation, with the difference that it does not contain a term related to the magnetic field:
  	\begin{equation}
	  	\frac{\partial \bm{\omega}_{\rm{n}}}{\partial t} = \bm{A}_{\omega \rm{n}}+\bm{B}_{\omega \rm{n}}+\bm{C}_{\omega \rm{n}}.
  	\end{equation}

\subsection{Comparison of induction and vorticity terms}
	By comparing the induction equation, Eq. (\ref{eq:induction2}), and the vorticity equation for the charged fluid, Eq. (\ref{eq:vort_c}), it can be checked that several of their terms share the dependence on the magnetic field, densities, pressures and velocities, but differ in their coefficients. The following relations are found:
	\begin{equation} \label{eq:bb_bc}
		\frac{|\bm{B}_{\rm{Ind}}|}{|\bm{B}_{\omega \rm{c}}|} = \frac{m_{\rm{i}}}{2 e} \sim 5 \times 10^{-9} \ \rm{kg \ C^{-1}},
	\end{equation}
	and
	\begin{equation} \label{eq:cb_cc}
		\frac{|\bm{C}_{\rm{Ind}}|}{|\bm{C}_{\omega \rm{c}}|} \sim \frac{m_{\rm{e}}}{e} \sim 5 \times 10^{-12} \ \rm{kg \ C^{-1}}.
	\end{equation}

	If we compare Eqs. (\ref{eq:bb_bc}) and (\ref{eq:cb_cc}), we get that
	\begin{equation}
		\frac{|\bm{C}_{\rm{Ind}}|}{|\bm{C}_{\omega \rm{c}}|} \ll \frac{|\bm{B}_{\rm{Ind}}|}{|\bm{B}_{\omega \rm{c}}|}.
	\end{equation} 
	The collisional term is much less important in the evolution of the magnetic field than in the evolution of vorticity in comparison with the battery and the baroclinic terms. This explains the results shown in Fig. \ref{fig:bfield_xc} for the series of simulations III (blue crosses): the effect of the collisional term in the induction equation is negligible.
	
	\begin{figure*} [t]
		\centering
		\includegraphics[width=0.329\hsize]{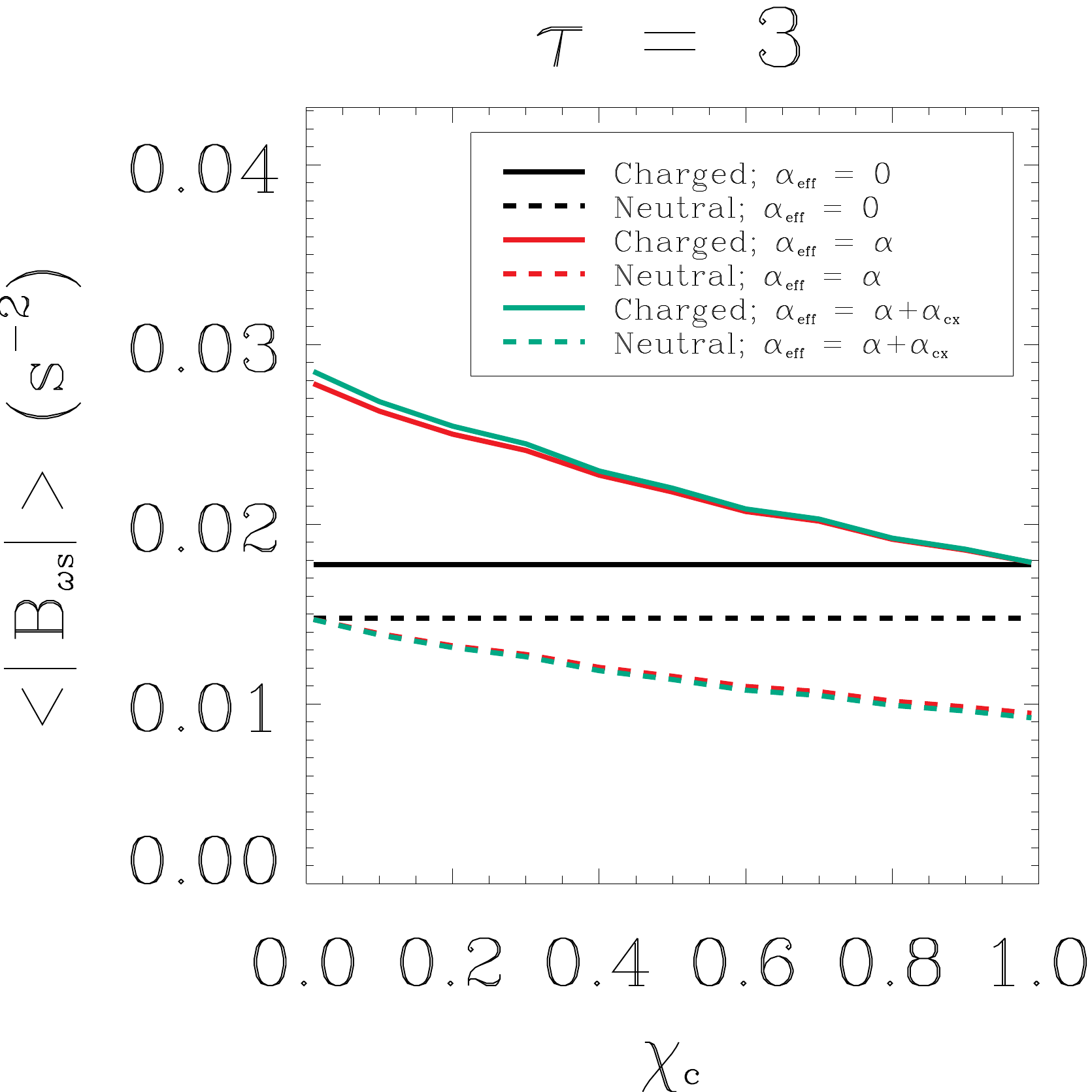}
		\includegraphics[width=0.329\hsize]{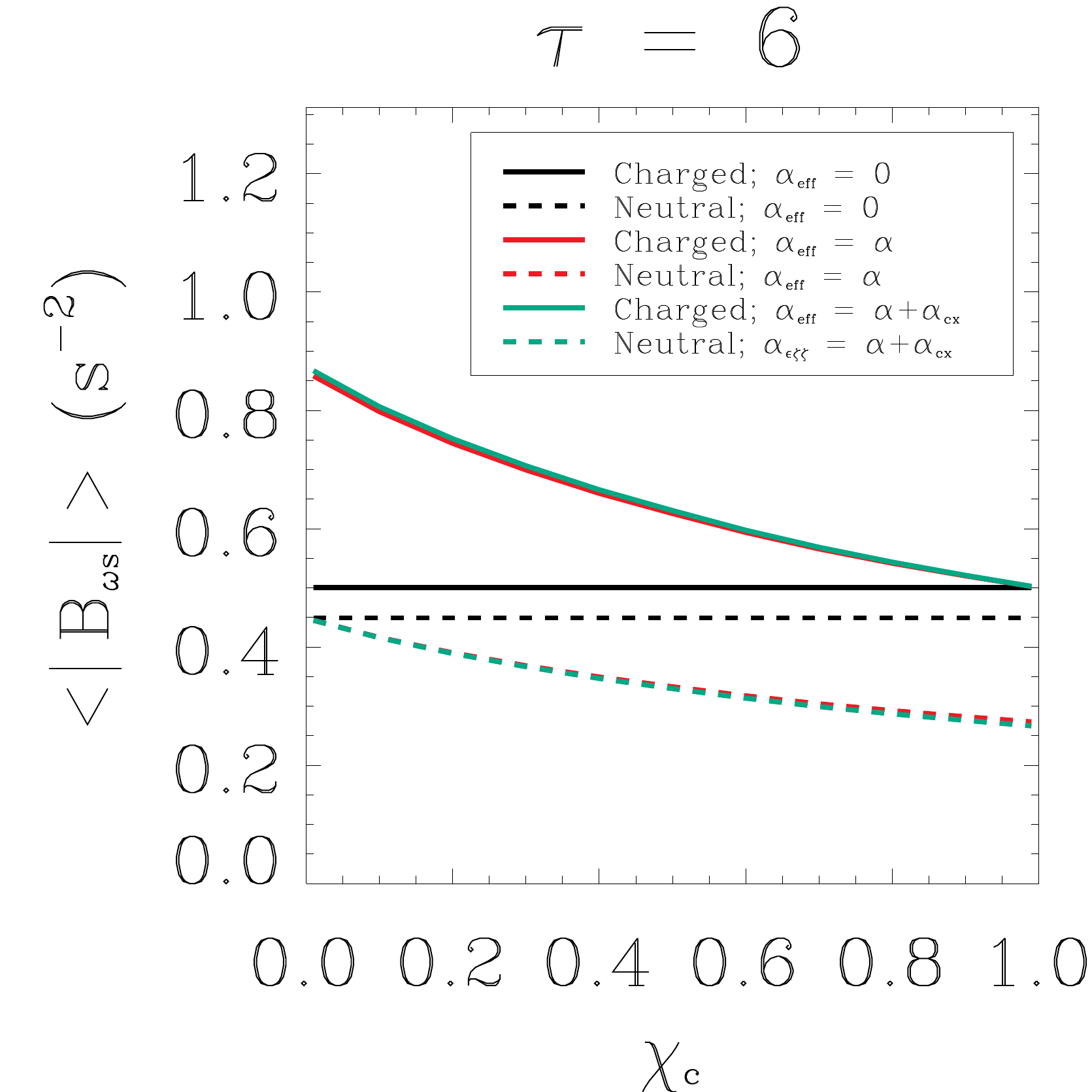}
		\includegraphics[width=0.329\hsize]{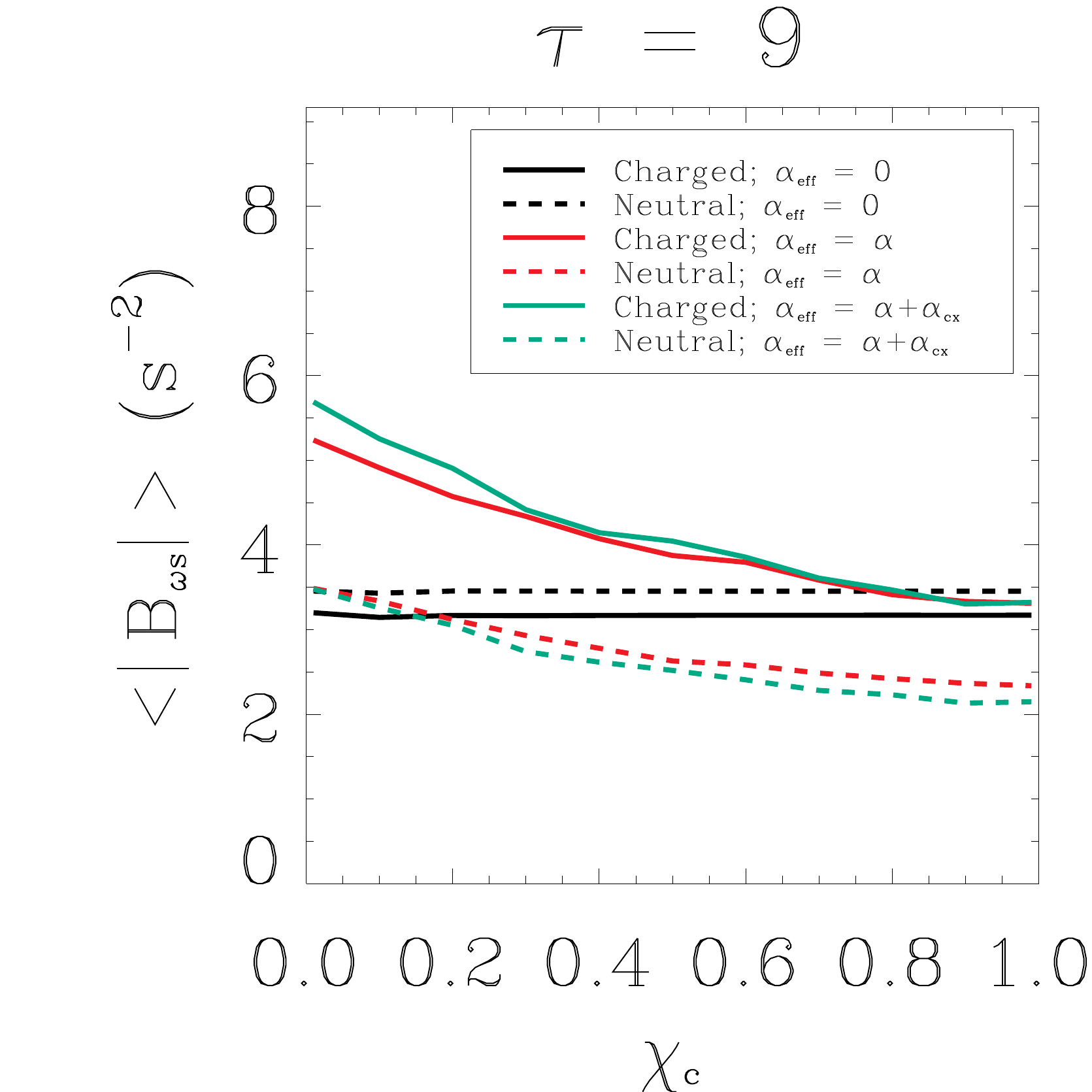}
		\caption{Spatially averaged values of the baroclinic terms as functions of the ionisation degree, $\chi_{\rm{c}}$, at three different times: $\tau = 3$ (left), $\tau = 6$ (middle) and $\tau = 9$ (right). Solid and dashed lines correspond to the terms $\langle|\bm{B}_{\omega \rm{c}}|\rangle$ and $\langle|\bm{B}_{\omega \rm{n}}|\rangle$, respectively. The black, red and green colours correspond to the cases with $\alpha_{\rm{eff}}=0$, $\alpha_{\rm{eff}}= \alpha$ and $\alpha_{\rm{eff}} = \alpha + \alpha_{\rm{cx}}$, respectively. }
		\label{fig:baroc_xc}
	\end{figure*}
	
	Now, performing a dimensional analysis of the baroclinic terms from the charged and neutral vorticity equations we get that
	\begin{equation} \label{eq:w_bs}
  		|\bm{B}_{\omega \rm{c}}| \sim \frac1{\gamma} \left(\frac{c_{\rm{S,c}}}{L_{\rm{ref}}} \right)^{2} \ \text{and} \ |\bm{B}_{\omega \rm{n}}| \sim \frac1{\gamma}\left(\frac{c_{\rm{S,n}}}{L_{\rm{ref}}} \right)^{2},
  	\end{equation}
  	where $L_{\rm{ref}}$ is a reference value of the length scales of the system. Furthermore, if we compare both terms and assume that the two fluids have the same temperature, we get the next relation:
  	\begin{equation} \label{eq:bc_bn_2}
		\frac{|\bm{B}_{\omega \rm{c}}|}{|\bm{B}_{\omega \rm{n}}|} \sim \frac{c_{\rm{S,c}}^{2}}{c_{\rm{S,n}}^{2}} = 2,
	\end{equation}
	which implies that the baroclinic term tends to produce a larger vorticity in the charged fluid than in the neutral one, due to the larger sound speed of the former. This relation, in combination with Eq. (\ref{eq:bb_bc}), is relevant for the explanation of the results from the simulations that we present in Sect. \ref{sec:eq_pt}.
	 
	Figure \ref{fig:baroc_xc} shows the average values of the baroclinic terms of the charged and neutral vorticity equations as functions of the ionisation degree at the times $\tau = 3$ (left), $\tau = 6$ (middle), and $\tau = 9$ (right). These results can be compared with Fig. \ref{fig:bfield_xc}. We remind that $|\bm{B}_{\rm{Ind}}| =|\bm{B}_{\omega \rm{c}}| m_{\rm{i}} / (2 e) $. Thus, the dependence depicted here for the charged baroclinic term is also applicable to the battery term.

	When there is no coupling between the two fluids, Fig. \ref{fig:baroc_xc} shows that the baroclinic terms do not change with the ionisation degree. This agrees with the prediction of Eq. (\ref{eq:batt_c}) for the battery term. However, when collisions are considered, each baroclinic term has a different behaviour: although both display a similar dependence on the ionisation degree (that is, their magnitudes decrease when $\chi_{\rm{c}}$ increases), $\bm{B}_{\omega \rm{c}}$ is larger than in the uncoupled case whereas $\bm{B}_{\omega \rm{n}}$ is smaller. The neutral baroclinic term (dashed lines) is less affected by the interaction with the charged fluid in the weakly ionised regime, while the charged baroclinic term (solid lines) is less affected by the interaction with the neutral fluid in the strongly ionised limit.
	
	Comparing Figs. \ref{fig:bfield_xc} and \ref{fig:baroc_xc} it can be seen that the magnetic field follows the same trends as the charged fluid baroclinic term or, equivalently, the battery term. However, this term does not directly depend on any kind of collisional parameter. So, the influence of the collisional interaction on the generation of magnetic field must come indirectly from the dependence of the Biermann battery term on the density or on the pressure of the charged fluid. The continuity equation that describes the temporal evolution of the density, Eq. (\ref{eq:rho}), does not involve any term related to collisions. On the other hand, the internal energy equations, Eqs. (\ref{eq:iener_c}) and (\ref{eq:iener_n}), do include source terms on their right-hand sides that represent the heat transfer due to the collisional interaction between the two fluids. We analyse those equations in the following section.
	
\subsection{Pressure, temperature and energy transfer} \label{sec:eq_pt}
	Inserting in Eqs. (\ref{eq:iener_c}) and (\ref{eq:iener_n}) the relations between the internal energies and the pressures given by Eq. (\ref{eq:iene_pres}), we obtain the following expressions for the temporal evolution of the charged and neutral pressures:
	\begin{equation} \label{eq:pres_c}
		\frac{\partial P_{\rm{c}}}{\partial t} + \bm{V}_{\rm{c}} \cdot \nabla P_{\rm{c}} + \gamma P_{\rm{c}} \nabla \cdot \bm{V}_{\rm{c}} = \left(\gamma - 1\right) \left[\bm{J} \cdot \bm{E}_{\rm{diff}} + Q_{\rm{cn}} \right]
	\end{equation}
	and
	\begin{equation} \label{eq:pres_n}
		\frac{\partial P_{\rm{n}}}{\partial t} + \bm{V}_{\rm{n}} \cdot \nabla P_{\rm{n}} + \gamma P_{\rm{n}} \nabla \cdot \bm{V}_{\rm{n}} = \left(\gamma -1\right) Q_{\rm{nc}},
	\end{equation}
	with $Q_{\rm{cn}}$ and $Q_{\rm{nc}}$ given by Eqs. (\ref{eq:qcn}) and (\ref{eq:qnc}), respectively. We remind here that $Q_{\rm{cn}} \ne Q_{\rm{nc}}$ but they follow the relation
	\begin{equation} \label{eq:qcn_qnc}
	    Q_{\rm{cn}} + Q_{\rm{nc}} = \alpha_{\rm{eff}} \rho_{\rm{c}} \rho_{\rm{n}} v_{\rm{D}}^{2},
	\end{equation}
	which is equivalent to the expression $Q_{\rm{ab}} + Q_{\rm{ba}} = -\bm{R}_{\rm{ab}} \cdot \left(\bm{V}_{\rm{a}}-\bm{V}_{\rm{b}}\right)$ from \citet{1965RvPP....1..205B}.
	
	For the present analysis, it is useful to express the energy transfer terms as $Q_{\rm{cn}} = Q_{\rm{cn}}^{T} + Q_{\rm{cn}}^{V}$ and $Q_{\rm{nc}} = Q_{\rm{nc}}^{T} + Q_{\rm{nc}}^{V}$, where
	\begin{equation}
	    Q_{\rm{cn}}^{T} = \frac{\alpha_{\rm{eff}}\rho_{\rm{c}}\rho_{\rm{n}}k_{\rm{B}}}{\left(\gamma - 1\right)m_{\rm{n}}}\left(T_{\rm{n}}-T_{\rm{c}}\right) = -Q_{\rm{nc}}^{T}
	\end{equation}
	are the terms depending on the difference of temperatures, and
	\begin{equation}
	    Q_{\rm{cn}}^{V} = Q_{\rm{nc}}^{V} = \alpha_{\rm{eff}}\rho_{\rm{c}}\rho_{\rm{n}}\frac{v_{\rm{D}}^{2}}{2}
	\end{equation}
	are the terms related to the velocity drifts. Since $Q_{\rm{cn}}^{V}$ and $Q_{\rm{nc}}^{V}$ are always positive, their effect is to increase the pressures (or internal energies) of both fluids, but the thermal exchange terms increase the pressure of one fluid while decreasing the pressure of the other one.
	
	Compared to the neutral pressure equation, the equation for the charges pressure includes Joule heating term that depends on the current density. However, due to the small values of the magnetic field present in the simulations, this term can be neglected in comparison with the collisional term, as shown in Appendix \ref{sec:app_joule}.
	
	Furthermore, according to the setup described in Sect. \ref{sec:setup}, the initial temperatures of both fluids are the same, $T_{\rm{c}} = T_{\rm{n}}$, and $Q_{\rm{cn}}^{T} = 0 = Q_{\rm{nc}}^{T}$ initially. To check if the thermal exchange terms keep negligible during the evolution as well, we focus now on the equations for the temperatures. 
	
	From Eqs. (\ref{eq:pres_c}) and (\ref{eq:pres_n}), with the assistance of the continuity equation, Eq. (\ref{eq:rho}), and the equation of state, Eq. (\ref{eq:state}), the following expression is obtained for each fluid $s \in \{c,n\}$:
	\begin{equation} \label{eq:temp}
		\frac{\partial T_{\rm{s}}}{\partial t} + \bm{V}_{\rm{s}} \cdot \nabla T_{\rm{s}} + \left(\gamma - 1\right)T_{\rm{s}} \nabla \cdot \bm{V}_{\rm{s}} = \left(\frac{\partial T_{\rm{s}}}{\partial t}\right)_{\rm{coll}} ,
	\end{equation}
	with
	\begin{equation} \label{eq:tcoll_c}
	    \left(\frac{\partial T_{\rm{c}}}{\partial t}\right)_{\rm{coll}} = \frac{\left(\gamma -1\right)}{n_{\rm{c}} k_{\rm{B}}} \left[Q_{\rm{cn}}^{T}+Q_{\rm{cn}}^{V}\right]
	\end{equation}
	and
	\begin{equation} \label{eq:tcoll_n}
	    \left(\frac{\partial T_{\rm{n}}}{\partial t}\right)_{\rm{coll}} = \frac{\left(\gamma - 1\right)}{n_{\rm{n}} k_{\rm{B}}}\left[-Q_{\rm{cn}}^{T}+Q_{\rm{cn}}^{V}\right].
	\end{equation}
	
	When there are no collisions, the temperature of each fluid evolves independently. Given that the velocities of neutrals and charges are not the same (see Fig. \ref{fig:vdiff}), it is expected that differences in the temperatures will appear during the evolution.
	
	The charged-neutral collisional interaction produces temporal variations of temperature that depend on the number density of each fluid, that is, they depend on the ionisation degree of the plasma. As a consequence of $Q_{\rm{cn}}^{V}$, the temperature of neutrals and charges will increase, but it will do in different amounts because the heating rate is inversely proportional to their corresponding number density. In contrast, $Q_{\rm{cn}}^{T}$ can increase or decrease the temperatures of neutrals and charges, depending on the sign of the difference of temperature between them. The equations for the temperature evolution under the influence of the solely $Q_{\rm{cn}}^{T}$ term show that the temperatures tend to equilibrate on the time scale given by 
	\begin{equation} \label{eq:tau_coll}
		\tau_{\rm{col}} = \frac{1}{\nu_{\rm{cn}}+\nu_{\rm{nc}}},
	\end{equation}
	as already shown by previous studies on partially ionised plasmas. Therefore, the net effect of the charged-neutral collisions is to increase the temperature of the two fluids by means of the dissipation of kinetic energy and then to lead the plasma to a new thermal equilibrium by means of the thermal exchange \citep{2005A&A...442.1091L,2011A&A...529A..82Z,2013ApJ...767..171S,2016ApJ...818..128O,2018ApJ...856...16M}.
	
	Coming back to the baroclinic terms in the vorticity equations, in the situation where $T_{\rm{c}}(x,z) \approx T_{\rm{n}}(x,z)$ (and consequently, $\nabla T_{\rm{c}} \approx \nabla T_{\rm{n}}$) holds, and also $\nabla \rho_{\rm{c}} / \rho_{\rm{c}} \approx \nabla \rho_{\rm{n}} / \rho_{\rm{n}}$ is assumed\footnote{We have checked that this assumption is supported by the results of the simulations.}, the relation $|\bm{B}_{\omega \rm{c}}| \sim 2 |\bm{B}_{\omega \rm{n}}|$ has to be valid for any ionisation degree, see  Eq. (\ref{eq:bc_bn_2}).
	According to Fig. \ref{fig:baroc_xc}, the neutral and charges baroclinic terms, $|\bm{B}_{\omega \rm{n}}|$ and   $|\bm{B}_{\omega \rm{c}}|$, do vary with ionisation degree $\chi_{\rm{c}}$. They both decrease when this parameter is increased. However, the ratio between both, $|\bm{B}_{\omega \rm{c}}|/|\bm{B}_{\omega \rm{n}}|$, keeps to be approximately 2, in agreement with our order of magnitude estimates above. In the weakly ionised limit,  $\chi_{\rm{c}} \rightarrow 0$, the values of $|\bm{B}_{\omega \rm{n}}|$ for the cases with and without collisions are almost equal. In the opposite limit of high ionisation,  $\chi_{\rm{c}} \rightarrow 1$, the charged baroclinic terms, $|\bm{B}_{\omega \rm{c}}|$, for the cases with and without collisions become equal.
	
	The reason for the dependence of the strength of the magnetic field generated by the battery term on the ionisation degree can now be understood taking into account the results from Fig. \ref{fig:baroc_xc} and the behaviour of the collisional terms in Eq. (\ref{eq:temp}). Consider in the first place a plasma that is almost fully ionised, with $n_{\rm{c}} \gg n_{\rm{n}}$. In that case, we find that
	\begin{equation}
		\left\lvert \left(\frac{\partial T_{\rm{c}}}{\partial t}\right)_{\rm{coll}} \right\rvert \ll \left\lvert \left(\frac{\partial T_{\rm{n}}}{\partial t}\right)_{\rm{coll}} \right\rvert,
	\end{equation}
	which means that the temperature of the charged fluid is almost unaffected by the interaction with the neutral fluid. The latter will be subject to much larger temporal variations in temperature. The baroclinic term of the charged fluid will remain almost the same as in the case without collisions. So, for the relation given by Eq. (\ref{eq:bc_bn_2}) to hold, $|\bm{B}_{\rm{n}}|$ is reduced.
	
	\begin{figure*}
		\centering
		\includegraphics[width=0.329\hsize]{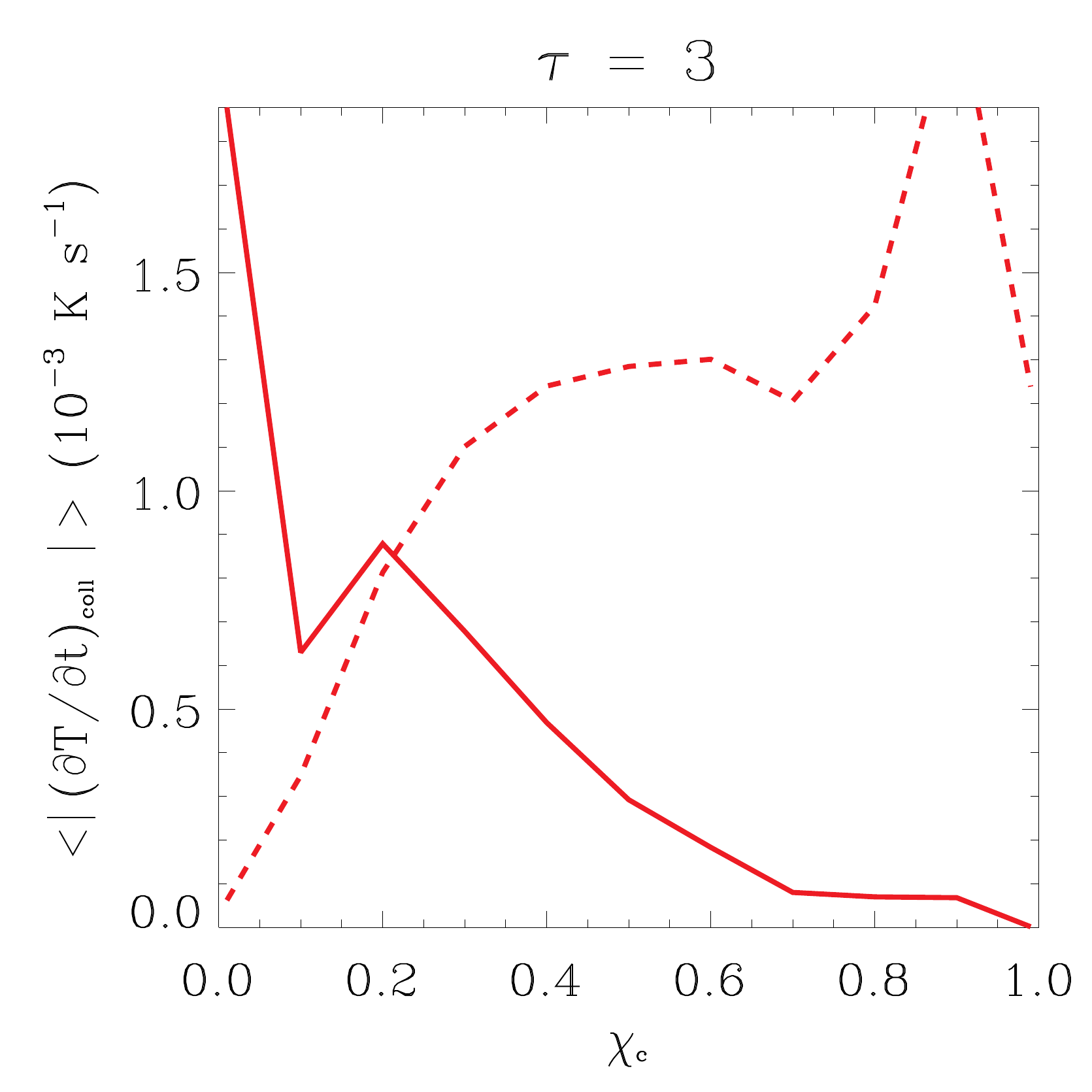}
		\includegraphics[width=0.329\hsize]{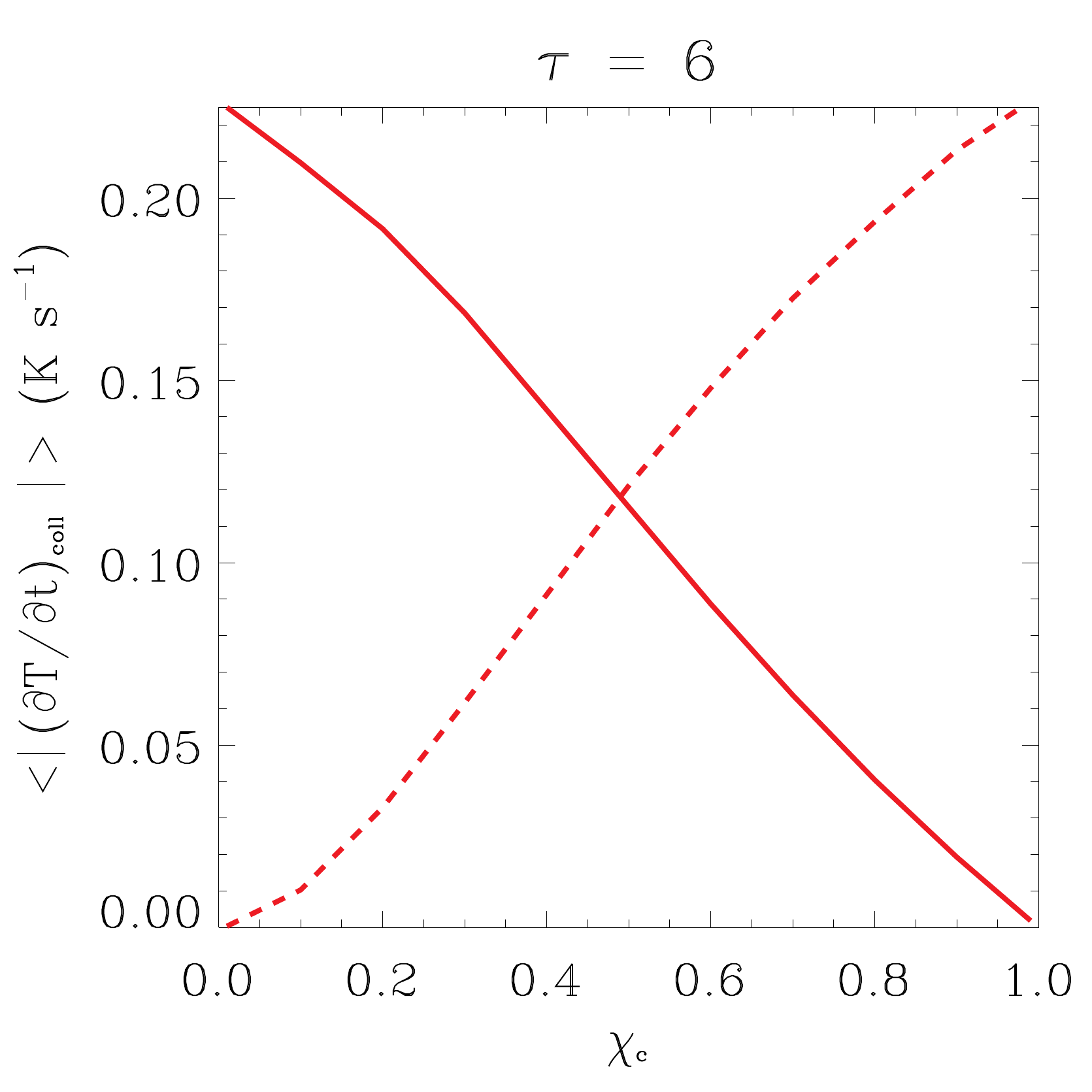}
		\includegraphics[width=0.329\hsize]{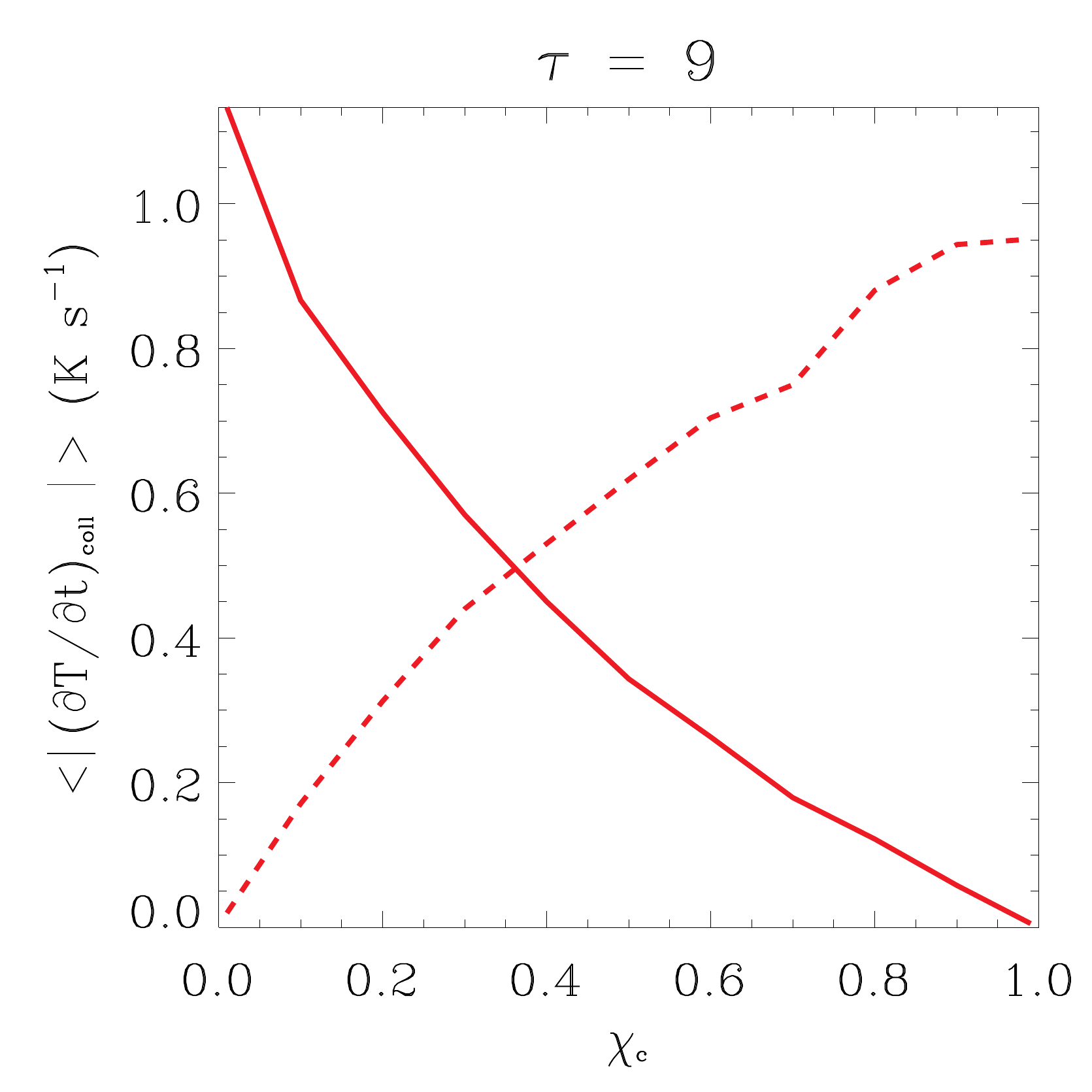}
		\caption{Spatially averaged values of the collisional terms in the temperature equations as functions of the ionisation degree, at three different times: $\tau = 3$ (left), $\tau = 6$ (middle) and $\tau = 9$ (right). Solid and dashed lines represent the terms for the charged and the neutral fluids, respectively, which are given by Eqs. (\ref{eq:tcoll_c}) and (\ref{eq:tcoll_n}).}
		\label{fig:tcolterm_xc}
	\end{figure*}
		
	On the other hand, in the weakly ionised regime, that is $n_{\rm{c}} \ll n_{\rm{n}}$, we get that
	\begin{equation}
		\left\lvert \left(\frac{\partial T_{c}}{\partial t}\right)_{\rm{coll}} \right\rvert \gg \left\lvert \left(\frac{\partial T_{n}}{\partial t}\right)_{\rm{coll}} \right\rvert,
	\end{equation}
	so now the charged fluid has much larger variations of temperature. This leads to an increase of the magnitude of the charged baroclinic term in comparison with the case without collisions. Consequently, the battery term, which is proportional to $|\bm{B}_{\omega \rm{c}}|$, also increases and produces stronger magnetic fields.
	
	The behaviour of the collisional terms described above can be checked in Fig. \ref{fig:tcolterm_xc}, where we have represented the average of their absolute values at different times of the simulations. By comparing Figs. \ref{fig:baroc_xc} and \ref{fig:tcolterm_xc} it can be seen that the baroclinic term for the charged fluid and the collisional term given by Eq. (\ref{eq:tcoll_c}) (represented by the solid red lines in the respective figures) follow similar trends: they increase as the ionisation degree of the plasma decreases.
	
\subsection{Comparison with single-fluid model}
	To study the origin of cosmic magnetic fields, \citet{1997ApJ...480..481K} used a single-fluid model to simulate the generation of magnetic fields by the Biermann battery term in partially ionised plasmas. With the notation of the present work, the induction equation of that single-fluid model can be written as
	
	\begin{equation} \label{eq:indu_single}
		\frac{\partial \bm{B}}{\partial t}=\nabla \times \left[\bm{V} \times \bm{B}\right] - \frac{m_{\rm{i}}}{e} \frac{1}{1+\chi_{\rm{c}}}\frac{\nabla \rho_{\rm{T}} \times \nabla P_{\rm{T}}}{\rho_{\rm{T}}^{2}}.
	\end{equation}
	Here, the variables $\bm{V}$ and $P_{\rm{T}}$ are the weighted mean velocity of the partially ionised plasma and the total pressure, and are given by
	\begin{equation}
		\bm{V} = \frac{\rho_{\rm{c}} \bm{V}_{\rm{c}}+\rho_{\rm{n}} \bm{V}_{\rm{n}}}{\rho_{\rm{T}}} = \chi_{\rm{c}} \bm{V}_{\rm{c}}+\left(1-\chi_{\rm{c}}\right)\bm{V}_{\rm{n}}
	\end{equation}
	and
	\begin{equation}
		P_{\rm{T}} = P_{\rm{c}} + P_{\rm{n}} = P_{\rm{e}} + P_{\rm{i}} + P_{\rm{n}}.
	\end{equation}
	
	The behaviour of the magnetic field as a function of the ionisation degree predicted by Eq. (\ref{eq:indu_single}) is similar to that depicted in Fig. \ref{fig:bfield_xc}. In the fully ionised limit ($\chi_{\rm{c}}=1$), Eq. (\ref{eq:indu_single}) is equivalent to the first line of Eq. (\ref{eq:induction2}). In addition, both the single-fluid model and the two-fluid model predict that when the ionisation degree is decreased, larger magnetic fields are generated. Nevertheless, the single-fluid model predicts a larger increase of the magnetic field. In the weakly ionised regime, when $\chi_{\rm{c}} \to 0$, the battery term in Eq. (\ref{eq:indu_single}) is a factor $\sim 2$ larger than in the fully ionised case. On the other hand, Fig. \ref{fig:bfield_xc} displays a maximum ratio between the two limits of $\sim 1.8$ when charge-exchange collisions are considered or $\sim 1.7$ when they are not.
	
	These discrepancies can be understood in terms of the assumptions that each model makes about the coupling degree between the two species. 
	\begin{figure}
		\centering
		\resizebox{\hsize}{!}{\includegraphics[]{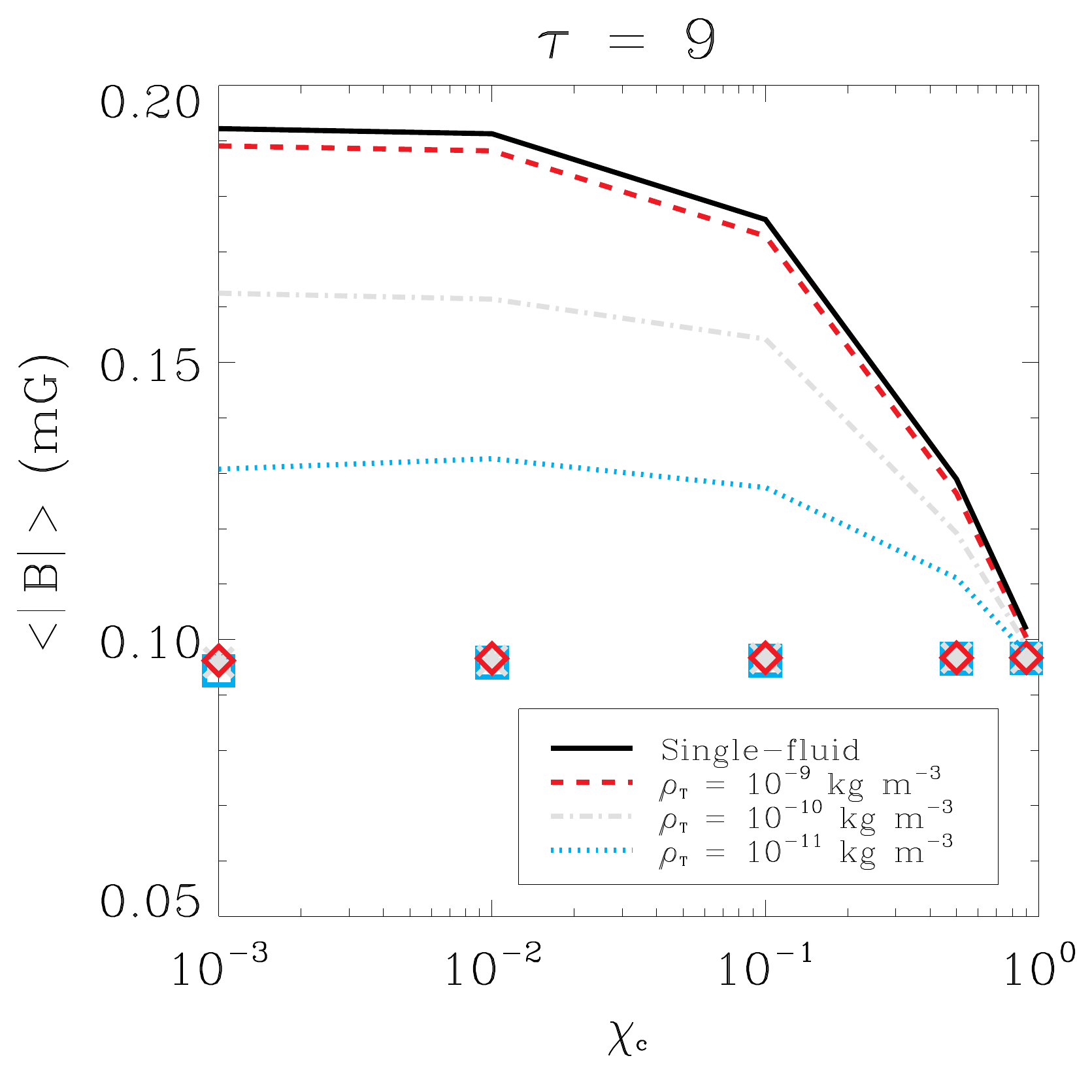}}
		\caption{Spatially averaged magnetic field as a function of ionisation degree for simulations with $N_{0} = 500$ and different values of total density, $\rho_{\rm{T}}$. Symbols and lines correspond to simulations with $\alpha_{\rm{eff}} = 0$ and $\alpha_{\rm{eff}} = \alpha$, respectively. The blue dotted, grey dashed-dotted and red dashed lines represent the cases with $\rho_{\rm{T}} = 10^{-11}, \ 10^{-10} \ \rm{and} \ 10^{-9} \ \rm{kg \ m^{-3}}$, respectively. The black line corresponds to the prediction of the single-fluid model.}		\label{fig:bfield_xc_rhos}
	\end{figure}

	The single-fluid model assumes a strong coupling between the two species, which have the same temperature and almost negligible velocity drifts, $v_{\rm{D}}$. In contrast, in the two-fluid model the strength of the coupling is not assumed \textit{a priori} but computed in each time step. It is represented by the factor $\alpha_{\rm{eff}} \rho_{\rm{c}} \rho_{\rm{n}}$, which greatly depends on the total density of the plasma and the ionisation degree, as clearly shown when expressed in the following way:
	\begin{equation} \label{eq:coupling}
		\alpha_{\rm{eff}} \rho_{\rm{c}} \rho_{\rm{n}} = \alpha_{\rm{eff}} \chi_{\rm{c}} \left(1-\chi_{\rm{c}}\right) \rho_{\rm{T}}^{2}.
	\end{equation} 
	
	For a given value of the total density, the largest friction coefficient is obtained when $\chi_{\rm{c}} = 0.5$, while it obviously tends to zero in the fully ionised and the fully neutral limits. Therefore, in the weakly ionised regime the coupling between the two species may be not as strong as it is assumed by the single-fluid model.
	
	To check how the coupling and, consequently, the generation of magnetic field depend on the total density, we have performed additional series of simulations with three different values of $\rho_{\rm{T}}$. The results of this study are presented in Fig. \ref{fig:bfield_xc_rhos}, in which the spatially averaged magnetic field is plotted as a function of the ionisation degree at the time $\tau = 9$. The blue dotted, grey dashed-dotted and red dashed lines represent the results for the cases with $\rho_{\rm{T}} = 10^{-11}, \ 10^{-10} \ \rm{and} \ 10^{-9}$, respectively, in which $\alpha_{\rm{eff}} = \alpha$. The symbols correspond to simulations with no charged-neutral collisions. The theoretical prediction of the single-fluid model is included as a solid black line and has been computed by multiplying the results with $\alpha_{\rm{eff}} = 0$ by a factor $2 / (1+\chi_{\rm{c}})$. It can be seen that the results from the two-fluid simulations approach the behaviour predicted by the single-fluid model as the total density of the plasma increases.
	
	It is also remarkable that the results presented in Fig. \ref{fig:bfield_xc_rhos} for the case with no collisions do not depend on the total density of the plasma. This is explained by the already mentioned fact that the evolution of the KHI does not depend on the total density but on the ratio of densities of the different media.
	
\section{Summary and future work} \label{sec:conc}
	In this paper we have investigated the effect that the partial ionisation in plasmas has on the Biermann battery mechanism. We have used the two-fluid model implemented in the numerical code \textsc{Mancha-2F} \citep{2019A&A...627A..25P} to simulate the evolution of the Kelvin-Helmholtz instability in plasmas with different ionisation degrees. From initial conditions with no magnetic field, we have studied how a weak magnetic field is created by means of the battery mechanism and how it evolves as the instability develops.
	
	First, we have have compared the evolution of the charged and the neutral fluids in the cases without and with coupling due to elastic collisions, for a particular value of the ionisation degree. We found that, when no collisions are considered,  the KHI develops slightly faster in the charged fluid than in the neutral one, although the initial conditions of density, velocity and temperature were identical for both of them. The explanation for this behaviour is that the charged and neutral fluids have different pressures (the charged fluid takes into account the pressure of electrons). As a consequence, the sound speed of the charged fluid is larger. With the density ratio between the internal and external regions of the plasma slabs chosen in our simulations, a larger sound speed leads to a slightly larger growth rate of the instability \citep{2012ApJ...749..163S}. In the case with collisions, the two fluids follow the same evolution at the large spatial scales but differences still appear at the smallest scales, in good agreement with the results of \citet{2019PhPl...26h2902H}.

	Then, we have found that the collisional interaction between the charged and neutral fluids enhances the generation of magnetic field by means of the Biermann battery mechanism (see Figs. \ref{fig:eb_resol} and \ref{fig:bfield_xc}). We have investigated how the generation and amplification of the magnetic field depends on the ionisation degree of the plasma and what is the role that different collisional mechanisms play in this processes.
	
	On the one hand, we have found that if there is no charged-neutral coupling, the generation of magnetic does not depend on the ionisation degree. The reason is that the evolution of the KHI does not depend on the total density of the charged fluid but on the density ratio between the different regions of the plasma, and this ratio has been kept the same for all the simulations.

	On the other hand, when collisions are taken into account, we have found that the generated magnetic field increases as the ionisation degree of the plasma decreases. The main reason for the increase of the magnetic field with decreasing ionisation degree comes from the elastic collisional terms included in the  momentum and energy equations. The induction equation also contains a term related to elastic collisions but its contribution is negligible. Finally, we have also considered the process of charge-exchange collisions and we have found that it slightly enhances the generated magnetic field, compared to the case with only elastic collisions.
	
	In order to find an explanation for the behaviour described above, we have made an analysis of the equations used in the two-fluid model. The Biermann battery term does not have any explicit dependence on the ionisation degree of the plasma or any parameter related to collisions. However, the simulations presented in this work show that the magnetic field generated by this mechanism does depend on the ionisation degree when the interaction between charges and neutrals is taken into account. Our analysis has shown that this dependence comes from the fact that the battery term is a function of the gradients of pressure or temperature and these variables are greatly affected by the presence of collisions.
	
	The charged-neutral interaction by means of collisions has two effects on the temperatures of the plasma. On the one hand, it increases the temperature of the two fluids by dissipating kinetic energy and transforming it into internal energy. The resulting increase of temperature is not the same for both fluids but depends on their number density, that is, on the ionisation degree. On the other hand, if the temperatures of the two fluids are different, collisions tend to reduce that difference until a thermal equilibrium is reached, with $T_{\rm{c}} \sim T_{\rm{n}}$. Under this equilibrium, the baroclinic terms from the vorticity equations tend to fulfill the relation $|\bm{B}_{\omega \rm{c}}| \sim 2 |\bm{B}_{\omega \rm{n}}|$, in contrast with the relation $|\bm{B}_{\omega \rm{c}}| \sim |\bm{B}_{\omega \rm{n}}|$ found when there is no coupling (as shown by Fig. \ref{fig:baroc_xc}). Collisions produce smaller variations of temperature in the fluid with a larger density. Therefore the magnitude of the baroclinic term of that fluid will have approximately the same value as in the case with no coupling. Then, in order for the relation $|\bm{B}_{\omega \rm{c}}| \sim 2|\bm{B}_{\omega \rm{n}}|$ to be fulfilled, the magnitude of the other baroclinic term is modified accordingly. This is why $|\bm{B_{\omega \rm{c}}}|$ increases as the ionisation degree decreases. And, since the Biermann battery term is proportional to the baroclinic term of the charged fluid, the generation of magnetic field is also enhanced as the ionisation degree is reduced. 
	
	Finally, we have compared our numerical results with the predictions from the single-fluid model used by \citet{1997ApJ...480..481K}. The single-fluid and the two-fluid model agree in the qualitative aspect: they both show that the generation of magnetic field by the Biermann battery mechanism is enhanced as the ionisation degree of the plasma is reduced. However, they differ in the magnitude of that enhancement. The single-fluid predicts a larger increase, which does not depend on the total density of the plasma. On the contrary, the results of the two-fluid model do depend on the total density because they depend on the coupling degree between the two species. This parameter is a function of the ionisation degree and the total density. As the total density is increased, the two-fluid results show a better agreement with the single-fluid predictions. The results presented in Fig. \ref{fig:bfield_xc_rhos} show that the trend of an increase of magnetic field as the ionisation degree decreases holds in the range $\chi_{\rm{c}} \in [10^{-3},1]$. However, according to Eq. (\ref{eq:coupling}), the coupling between the two fluids should tend to zero as the ionisation degree tends to zero. And without coupling there should not be enhancement of the Biermann mechanism. For future works, it would be interesting to extend the present study to even lower values of ionisation degree and check whether the mentioned trend still holds.
	
	For the simulations analysed in the present work, we have chosen values of density, temperatures, velocities and length-scales that correspond to plasmas in the solar chromosphere. However, our results may be easily extrapolated and be relevant to other astrophysical scenarios in which partially ionised plasmas are involved. The time and length scales of the evolution of the KHI and the strength of the magnetic field generated by the battery term would be different, but its qualitative behaviour regarding the interaction between the charged and neutral components of the plasma would be the same.
	
	There are some results from our simulations that we have briefly mentioned but left unexplored in more depth, such as certain trends observed in the velocity drifts when the collisional coupling is considered (see Fig. \ref{fig:vdiff}). This issue should be analysed in more detail in future works.
	
	Furthermore, here we have assumed that the ionisation degree is uniform while, in general, it should be a function of the temperature and density of the plasma. In addition, it would be interesting to study the dependence of the results on the initial conditions of temperature. A larger temperature corresponds to a larger value of the sound speed, which brings the plasma closer to the incompressible regime. With the setup chosen for our simulations, and according to the results of \citet{2012ApJ...749..163S}, this would produce larger growth rates of the KHI. Therefore, stronger magnetic fields would be generated during the first stages of the instability. However, a larger growth rate also means that the dissipation scales are reached faster and the maximum magnetic field obtained may be smaller than in a case with a lower temperature.
	
	Another important step would be to perform 3D simulations. The values of the magnetic field obtained in our 2D simulations are very small in comparison to those that are found in reality. It is expected that a 3D configuration will lead to a considerably larger amplification of the magnetic field, since it will not be restricted to only one spatial direction as in the 2D simulations of the present work. In addition, some relevant physical processes that have been neglected here, such as the real viscosity of the fluids or the thermal conduction, should be taken into account in the future, which would provide more realistic results.

\begin{acknowledgements}
	This work was supported by the European Research Council through the Consolidator Grant ERC-2017-CoG-771310-PI2FA. We thankfully acknowledge the technical expertise and assistance provided by the Spanish Supercomputing Network (Red Espanola de Supercomputacion), as well as the computer resources used: the LaPalma Supercomputer, located at the Instituto de Astrofisica de Canarias. We also thank the anonymous referee for constructive remarks and helpful suggestions.
\end{acknowledgements}

\bibliographystyle{aa}
\bibliography{battery2f_bib}

\appendix
\section{Effect of numerical dissipation} \label{sec:sim_diss}	
	In the \textsc{Mancha-2F} code, the numerical diffusion terms are implemented in such a way that they replicate the effects of viscosity and resistivity. This means that, for instance, the following term is added to the right-hand side of the momentum equation, Eq. (\ref{eq:momc}):
	\begin{equation}
	    \left(\frac{\partial \left(\rho_{\rm{c}}\bm{V}_{\rm{c}}\right)}{\partial t}\right)_{\rm{art}} = \nabla \cdot \hat{\tau}_{\rm{c,art}},
	\end{equation}
	where $\hat{\tau}_{\rm{c,art}}$ is the numerical viscous stress tensor. The components of this tensor are computed as follows:
	\begin{equation}
	    \hat{\tau}_{\rm{c,art,kl}} = \frac{1}{2} \rho_{\rm{c}}\left(\nu_{k} \frac{\partial V_{\rm{c,l}}}{\partial x_{k}} + \nu_{l}\frac{\partial V_{\rm{c,k}}}{\partial x_{l}} \right),
	\end{equation}
	with
	\begin{equation} \label{eq:nu_l}
	    \nu_{l} = a_{\rm{diff}} c_{\rm{S,c}} \Delta x_{l}.
	\end{equation}
	Here, $a_{\rm{diff}}$ is a constant that controls the strength of the numerical diffusion, $c_{\rm{S,c}}$ is the sound speed of the charged fluid and $\Delta x_{l}$ is the resolution of the numerical domain.
	
	On the other hand, to take into account the effect of the physical viscosity, the right-hand side of the momentum equation for each species $s$ should include the term
	\begin{equation}
	    \left(\frac{\partial \left(\rho_{\rm{s}}\bm{V}_{\rm{s}}\right)}{\partial t}\right)_{\rm{visc}} = \nabla \cdot \hat{\tau}_{\rm{s}},
	\end{equation}
	where
	\begin{equation}
	    \tau_{\rm{s},ij} = \xi_{\rm{s}} \left(\frac{\partial V_{\rm{s},i}}{\partial x_{j}} + \frac{\partial V_{\rm{s},j}}{\partial x_{i}}-\frac{2}{3}\delta_{ij} \nabla \cdot \bm{V}_{\rm{s}}\right)
	\end{equation}
	and the viscosity $\xi_{\rm{s}}$ is computed as \citep{1965RvPP....1..205B,2013PhPl...20f1202L}
	\begin{equation} \label{eq:visc_s}
	    \xi_{\rm{s}} = \frac{n_{\rm{s}}k_{\rm{B}}T_{\rm{s}}}{\nu_{\rm{ss}}} = \frac{\sqrt{\pi k_{\rm{B}}T_{\rm{s}}m_{\rm{s}}}}{2 \sigma_{\rm{ss}}},
	\end{equation}
	where the definition of the collision frequency $\nu_{\rm{ss}}$ given by Eq. (\ref{eq:nu_st}) has been used and $\sigma_{\rm{ss}}$ represents the cross-section of collisions between particles of the same species.
	
	\begin{figure} [!h]
		\centering
		\resizebox{0.92\hsize}{!}{\includegraphics[]{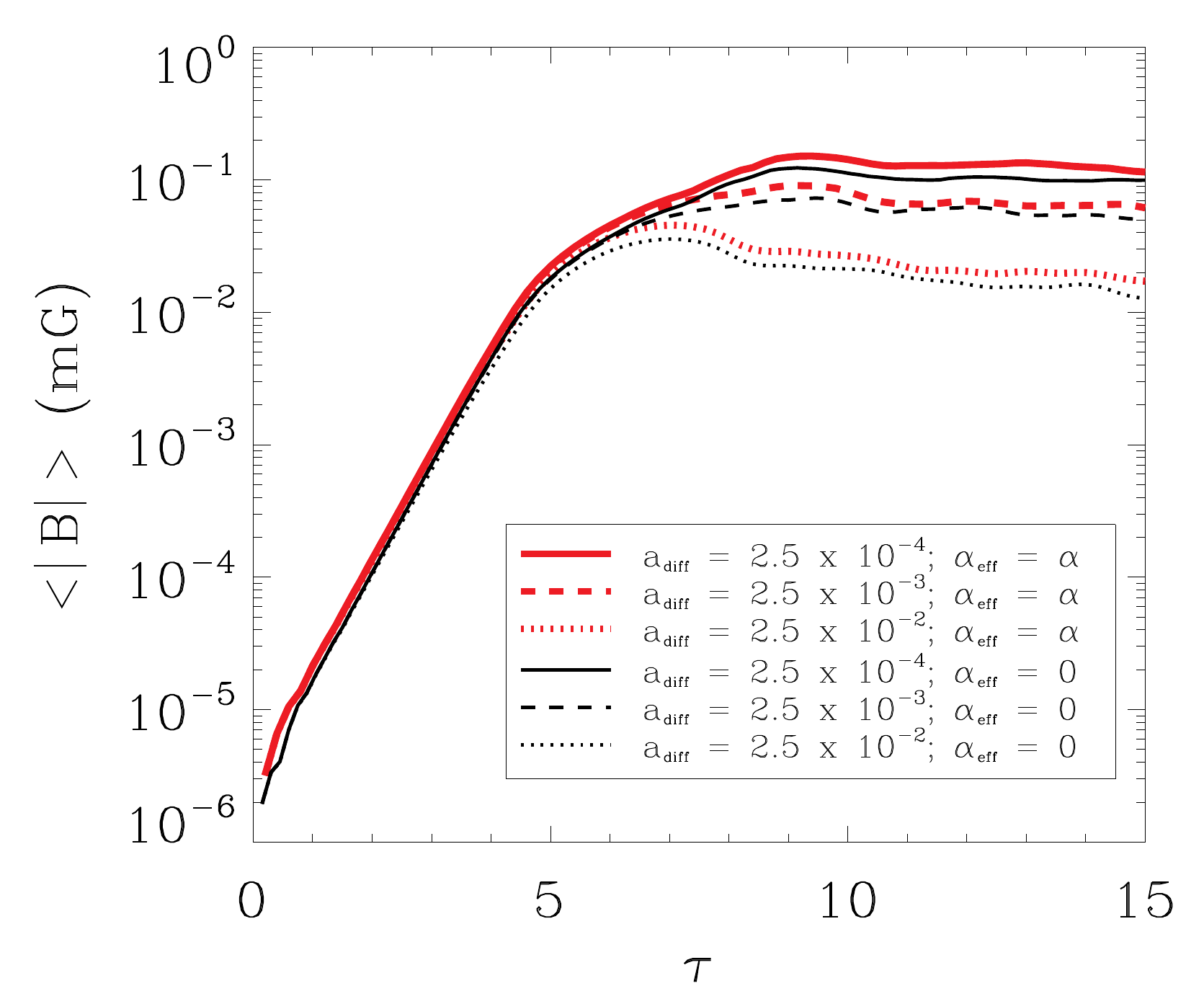}}
		\caption{Mean magnetic field as a function of time and numerical diffusivity for the case with $\chi_{\rm{c}} = 0.5$. The solid line represents a simulation with $a_{\rm{diff}} = 2.5 \times 10^{-4}$ (in normalised units) while the dashed and the dotted lines correspond to simulations with $a_{\rm{diff}} = 2.5 \times 10^{-3}$ and $a_{\rm{diff}} = 2.5 \times 10^{-2}$, respectively. Black and red lines represent the cases with $\alpha_{\rm{eff}} = 0$ and $\alpha_{\rm{eff}} = \alpha$.}
		\label{fig:eb_diffu}
	\end{figure}
	
	With the above definitions, we can compute the Reynolds numbers. Using the charged fluid as reference, the former dimensionless number is given by
	\begin{equation} \label{eq:re_real}
	    R_{e} = \frac{\nabla \cdot \left(\rho_{\rm{c}} \bm{V}_{\rm{c}} \bm{V}_{\rm{c}}\right)}{\nabla \cdot \hat{\tau}_{\rm{c}}} \sim \frac{\rho_{\rm{c}}V_{\rm{c}}^{2} / L}{\xi_{\rm{c}} V_{\rm{c}}/L^{2}} = \frac{\rho_{\rm{c}}V_{\rm{c}}L}{\xi_{\rm{c}}},
	\end{equation}
	where $V_{\rm{c}}$ and $L$ are reference values of the velocity of the flow and the length-scale of the system, respectively. In addition, the density of the charged fluid can be expressed in terms of $\rho_{T} = \rho_{\rm{c}} + \rho_{\rm{n}}$, the total density of the plasma, as $\rho_{\rm{c}} = \chi_{\rm{c}} \rho_{T}$, where $\chi_{\rm{c}}$ is the ionisation degree. Then, the Reynolds number can be rewritten as
	\begin{equation} \label{eq:re_real_2}
	    R_{e} = \frac{\chi_{\rm{c}} \rho_{T} V_{\rm{c}}L}{\xi_{\rm{c}}}.
	\end{equation}
	Therefore, the Reynolds number varies as the ionisation degree of the plasma is varied.
	
	For the purposes of this analysis, we can define a dimensionless number which plays an analogue role to Reynolds number. Here we refer to that quantity as the numerical Reynolds number, which is given by
	\begin{equation} \label{eq:re_art}
	    R_{e,\rm{num}} = \frac{\nabla \cdot \left(\rho_{\rm{c}} \bm{V}_{\rm{c}} \bm{V}_{\rm{c}}\right)}{\nabla \cdot \hat{\tau}_{\rm{c,art}}} \sim \frac{V_{\rm{c}} L}{\nu_{l}}
	\end{equation}
	and does not depend on the ionisation degree, in contrast with the real Reynolds number.
	
	For our simulations, we have decided to neglect the effect of the physical viscosity and only take into account the numerical one. In this way, the effective Reynolds number is kept the same for every series of simulations and the comparisons between each other are more straightforward.
    \begin{figure*}
		\centering
		\includegraphics[width=17cm]{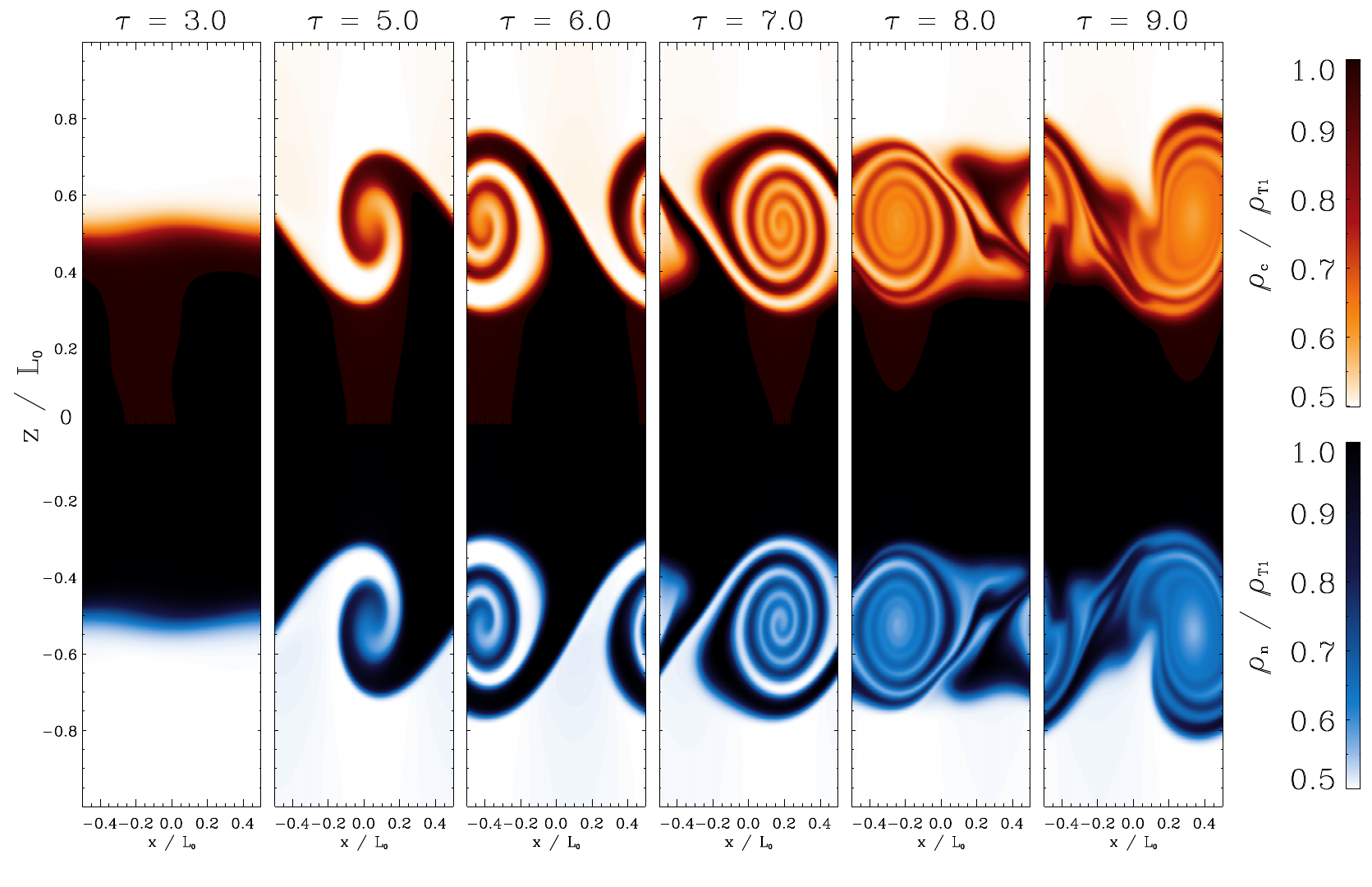}
		\caption{Density snapshots of a simulation with $\chi_{\rm{c}} = 0.5$, $\alpha_{\rm{eff}} = \alpha$, and a high value of numerical diffusivity, $a_{\rm{diff}} = 2.5 \times 10^{-2}$. Same colour scheme as in Fig. \ref{fig:rhos} is used. An animation of this figure is available online.}
		\label{fig:rho_diffu}
	\end{figure*}
	
	As described in \citet{2005A&A...429..335V} and \citet{2010ApJ...719..357F}, the strength of the numerical dissipation applied to each one of the temporal evolution equations presented in Sect. \ref{sec:eqs} is controlled by a different coefficient. However, in this work we have chosen the same value for all the coefficients, given by $a_{\rm{diff}}$.
	
	In Fig. \ref{fig:eb_diffu} we present a comparison of the effect of numerical diffusivity for the cases of $\chi_{\rm{c}} = 0.5$ with $\alpha_{\rm{eff}} = 0$ (black lines) and $\alpha_{\rm{eff}} = \alpha$ (red lines), and $N_{0} = 1000$. The spatially averaged magnetic field is plotted as a function of the normalised time. The results for the three values of $a_{\rm{diff}}$ coincide almost perfectly during the linear phase of the instability, while huge differences appear at later stages. For example, at the end of the simulation ($\tau = 15$), the magnetic field for the case with $a_{\rm{diff}} = 2.5 \times 10^{-4}$ is approximately one order of magnitude larger than for the case with the strongest diffusivity. The behaviour of the magnetic field displayed by this figure supports the statement that most of the dissipation occurs at the smallest length scales.
	
	The effect of a larger numerical diffusivity can also be checked in Fig. \ref{fig:rho_diffu}, which displays density snapshots for the case with $a_{\rm{diff}} = 2.5 \times 10^{-2}$. The comparison of Figs. \ref{fig:rhos} and \ref{fig:rho_diffu} shows that in the former the edges of the vortexes are sharper while in the latter they are more blurred. Furthermore, the secondary vortexes present at $\tau = 7$ in the case of low diffusivity do not appear when a much larger diffusivity is used. Finally, most of the small scale structure that can be seen in the right panels of Fig. \ref{fig:rhos} is removed in Fig. \ref{fig:rho_diffu}.
	
\section{Additional parameters of the plasma}
	Using the definition given by Eq. (\ref{eq:visc_s}), the physical viscosities of the charged and neutral fluid are $\xi_{\rm{c}} \approx 1.2 \times 10^{-7} \ \rm{kg \ m^{-1} \ s^{-1}} \ \text{and} \ \xi_{\rm{n}} \approx 1.7 \times 10^{-5} \ \rm{kg \ m^{-1} \ s^{-1}}$, respectively,
	where the following collisional cross sections have been used: $\sigma_{\rm{cc}} = \sigma_{\rm{ii}} \approx 1.2 \times 10^{-16} \ \rm{m^{2}}$ and $\sigma_{\rm{nn}} = 7.73 \times 10^{-19} \ \rm{m^{2}}$ \citep{2012PhPl...19g2508M,2013PhPl...20f1202L}.
	
	The	thermal conductivities, $\kappa_{\rm{c}}$ and $\kappa_{\rm{n}}$, are given by (\citet{1965RvPP....1..205B,2013PhPl...20f1202L}:
	\begin{equation} \label{eq:conduc_c}
	    \kappa_{\rm{c}} = \frac{2k_{\rm{B}}}{\sigma_{\rm{cc}}}\sqrt{\frac{\pi k_{\rm{B}}T_{\rm{c}}}{m_{\rm{c}}}} \approx 3.8 \times 10^{-3} \ \rm{W \ m^{-1} \ K^{-1}},
	\end{equation}
	and
	\begin{equation} \label{eq:conduc_n}
	    \kappa_{\rm{n}} = \frac{2k_{\rm{B}}}{\sigma_{\rm{nn}}}\sqrt{\frac{\pi k_{\rm{B}}T_{\rm{n}}}{m_{\rm{n}}}}\approx 0.57 \ \rm{W \ m^{-1} \ K^{-1}}.
	\end{equation}
    
    Then, the Peclet numbers, which represent the ratio of advective transport rate to the diffusive transport rate due to thermal conduction, are given by
	\begin{equation}
	    P_{e,\rm{c}} = \frac{L_{0}U_{0}\rho_{\rm{c}}c_{\rm{p,c}}}{\kappa_{\rm{c}}} \approx 13600
	\end{equation}
	and
	\begin{equation}
	    P_{e,\rm{n}} = \frac{L_{0}U_{0}\rho_{\rm{n}}c_{\rm{p,n}}}{\kappa_{\rm{c}}} \approx 100,
	\end{equation}
	where $c_{\rm{c,s}} = \gamma/(\gamma-1) k_{\rm{B}}/m_{s}$ is the heat capacity of species $s$. The values of these dimensionless numbers reveal that the thermal conduction, specially that of the neutral fluid, may play a relevant role in the dynamics of the plasma. However, for the sake of simplicity and to focus only on the study of the charged-neutral collisional interaction, we have not taken into account the effect of thermal conduction in the present research.
    
\section{Comparison between Joule and collisional heating} \label{sec:app_joule}
    Taking into account the definition for the variable $\bm{E}_{\rm{diff}}$, given by Eq. (\ref{eq:ediff}), and Ohm's law, Eq. (\ref{eq:ohm}), we get that the Joule heating is given by
	\begin{equation}
		\bm{J} \cdot \bm{E}_{\rm{diff}} = \bm{J} \cdot \left[-\eta_{\rm{H}} \nabla P_{\rm{e}} +\eta_{\rm{D}} \left(\bm{V}_{\rm{n}}-\bm{V}_{\rm{c}}\right)\right].
	\end{equation}
	The results presented in the main text demonstrate that the collisional term of Ohm's law can be neglected. So, the previous equation is reduced to
	\begin{equation}
		\bm{J} \cdot \bm{E}_{\rm{diff}} = - \eta_{\rm{H}} \bm{J} \cdot \nabla P_{\rm{e}} = - \frac{1}{e n_{\rm{e}}}\frac{\nabla \times \bm{B}}{\mu_{0}} \cdot \nabla P_{\rm{e}}.
	\end{equation}
	
	Then, the ratio between the Joule heating term and the collisional term ($Q_{\rm{cn}}^{T}$) can be estimated as:
	\begin{equation}
		\frac{|\bm{J} \cdot \bm{E}_{\rm{diff}}|}{|Q_{\rm{cn}}^{T}|} \sim \frac{\left(\gamma - 1\right) B_{\rm{ref}}}{\left(\Delta x\right)^{2}e \mu_{0} n_{\rm{c}} \nu_{\rm{cn}}} \sim 10^{-4},
	\end{equation}
	where the value of $B_{\rm{ref}}$ corresponds to the maximum magnetic field found in the simulations. This relation demonstrates that the term $\bm{J} \cdot \bm{E}_{\rm{diff}}$ can be neglected in the present study.

\end{document}